\documentclass[aps,prb,showpacs,twocolumn,amsmath,amssymb,superscriptaddress]{revtex4-1}
\usepackage[utf8]{inputenc}
\usepackage{booktabs}
\usepackage[multiple]{footmisc}
\usepackage{dcolumn}
\usepackage{rotating}
\usepackage{perpage}
\usepackage{xcolor}
\usepackage{braket} 
\usepackage{soul}
\usepackage{tikz}
\usepackage[T1]{fontenc}
\usepackage{etoolbox}
\usepackage{graphics}
\usepackage{siunitx}
\usepackage{hyperref}
\usepackage{float}
\usepackage{multirow}
\usepackage{mathtools}
\usepackage{bm}
\usepackage{url}
\usepackage{amsmath}
\usepackage{caption}
\usepackage{subfig}

\makeatletter
\def\upintkern@{\mkern-7mu\mathchoice{\mkern-3.5mu}{}{}{}}
\def\upintdots@{\mathchoice{\mkern-4mu\@cdots\mkern-4mu}%
 {{\cdotp}\mkern1.5mu{\cdotp}\mkern1.5mu{\cdotp}}%
 {{\cdotp}\mkern1mu{\cdotp}\mkern1mu{\cdotp}}%
 {{\cdotp}\mkern1mu{\cdotp}\mkern1mu{\cdotp}}}

\newcommand{\UpMultiIntegral}[1]{%
  \edef\ints@c{\noexpand\upintop
    \ifnum#1=\z@\noexpand\upintdots@\else\noexpand\upintkern@\fi
    \ifnum#1>\tw@\noexpand\upintop\noexpand\upintkern@\fi
    \ifnum#1>\thr@@\noexpand\upintop\noexpand\upintkern@\fi
    \noexpand\upintop
    \noexpand\ilimits@
  }%
  \futurelet\@let@token\ints@a
}
\makeatother

\DeclareFontFamily{OMX}{mdbch}{}
\DeclareFontShape{OMX}{mdbch}{m}{n}{ <->s * [0.8]  mdbchr7v }{}
\DeclareFontShape{OMX}{mdbch}{b}{n}{ <->s * [0.8]  mdbchb7v }{}
\DeclareFontShape{OMX}{mdbch}{bx}{n}{<->ssub * mdbch/b/n}{}

\DeclareSymbolFont{uplargesymbols}{OMX}{mdbch}{m}{n}
\SetSymbolFont{uplargesymbols}{bold}{OMX}{mdbch}{b}{n}
\DeclareMathSymbol{\upintop}{\mathop}{uplargesymbols}{82}
\DeclareMathSymbol{\upointop}{\mathop}{uplargesymbols}{"48}

\DeclareFontEncoding{MDB}{}{}
\DeclareFontFamily{MDB}{mdbch}{}
\DeclareFontShape{MDB}{mdbch}{m}{n}{ <->s * [0.8]  mdbchrmb }{}
\DeclareFontShape{MDB}{mdbch}{b}{n}{ <->s * [0.8]  mdbchbmb }{}
\DeclareFontShape{MDB}{mdbch}{bx}{n}{<->ssub * mdbch/b/n}{}
\DeclareFontSubstitution{MDB}{cmr}{m}{n}
\DeclareSymbolFont{mathdesignB}{MDB}{mdbch}{m}{n}%
\SetSymbolFont{mathdesignB}{bold}{MDB}{mdbch}{b}{n}%
\DeclareMathSymbol{\upintclockwise}{\mathop}{mathdesignB}{128}
\DeclareMathSymbol{\upointclockwise}{\mathop}{mathdesignB}{130}
\DeclareMathSymbol{\upointctrclockwise}{\mathop}{mathdesignB}{132}
\DeclareMathSymbol{\upoiint}{\mathop}{mathdesignB}{134}
\DeclareMathSymbol{\upoiiint}{\mathop}{mathdesignB}{136}

\makeatletter
\newcommand{\upint}{\DOTSI\upintop\ilimits@}
\newcommand{\upoint}{\DOTSI\upointop\ilimits@}
\makeatother

\begin{document}
\title{Efficient spin filtering through Fe$_4$GeTe$_2$-based van der Waals heterostructures}

\author{Masoumeh Davoudiniya}
\affiliation{Department of Physics and Astronomy, Uppsala University, Box 516,
 751\,20 Uppsala, Sweden}
 \author{Biplab Sanyal}
\email{For correspondence: Biplab.Sanyal@physics.uu.se}
\affiliation{Department of Physics and Astronomy, Uppsala University, Box 516, 751\,20 Uppsala, Sweden}

\date{\today}

\begin{abstract}
Utilizing ab initio simulations, we study the spin-dependent electronic transport characteristics within Fe$_4$GeTe$_2$-based van der Waals heterostructures. The electronic density of states for both free-standing and device-configured Fe$_4$GeTe$_2$ (F4GT) confirms its ferromagnetic metallic nature and reveals a weak interface interaction between F4GT and PtTe$_2$ electrodes, enabling efficient spin filtering. We observe a decrease in the magnetic anisotropy energy of F4GT in the device configuration, indicating reduced stability of magnetic moments and heightened sensitivity to external conditions. The transmission eigenstates of PtTe$_2$/ monolayer F4GT/PtTe$_2$ heterostructures demonstrate interference patterns affected by relative phases and localization, notably different in the spin-up and spin-down channels. The ballistic transport through a double-layer F4GT with a ferromagnetic configuration sandwiched between two PtTe$_2$ electrodes is predicted to exhibit an impressive spin polarization of 97$\%$ with spin-up electrons exhibiting higher transmission probability than spin-down electrons. 
Moreover, we investigate the spin transport properties of Fe$_4$GeTe$_2$/GaTe/Fe$_4$GeTe$_2$ van der Waals heterostructures sandwiched between PtTe$_2$ electrodes to explore their potential as magnetic tunnel junctions (MTJs) in spintronic devices. 
The inclusion of GaTe as a 2D semiconducting spacer between F4GT layers results in a tunnel magnetoresistance (TMR) of 487$\%$ at low bias and decreases with increasing bias voltage. In general, our findings underscore the potential of F4GT / GaTe / F4GT heterostructures to advance spintronic devices based on van der Waals materials.

\end{abstract}
\maketitle
\section{Introduction}
Quantum transport through two-dimensional (2D) magnetic structures has emerged as a fascinating field of research with promising implications for spintronics~\cite{AWSCHALOM1999130, Su2021, Li2021, lin, Li20212}. 
Spintronics aims to harness both the charge and spin degrees of freedom to enable the development of novel electronic devices with enhanced functionalities. Moreover, 2D magnetic structures, such as atomically thin ferromagnetic films and magnetic heterostructures, possess distinct spin-dependent properties that enable efficient control and manipulation of electron spins. Understanding spin transport in these systems is crucial for the development of spin-based electronic devices and spintronic circuits. Historically, most magnetic tunnel junctions (MTJs) were constructed using perovskite-oxide materials. However, these conventional MTJs have limitations, notably a large resistance-area product, which restricts their practicality in device applications~\cite{Velev2009}. In contrast, van der Waals (vdW) materials have shown promise in overcoming the challenges associated with traditional magnetic thin films. They lead to significantly high tunnel magnetoresistance (TMR) values, as evidenced by numerous experimental studies~\cite{10.1063/1.4930311, Karpiak_2020, Zhu2023, Huang2023, doi:10.1126/science.aar4851, doi:10.1126/science.aar3617}.
The vdW  heterostructures, composed of atomically thin layers stacked on top of each other, have emerged as promising platforms to explore and exploit such spin-dependent phenomena~\cite{Su2021, Devaraj, Kurebayashi2022, Sierra2021, Wang2018, PhysRevB.93.014411, PhysRevApplied.16.034052}. The weak vdW forces facilitate the formation of a clean and atomically sharp interface between layers, enabling efficient transfer of spin-polarized electrons between the magnetic materials.

\begin{figure*}[t]
\centering
\includegraphics[width=0.25\linewidth, trim={0cm 0cm 0cm 0cm}, clip]{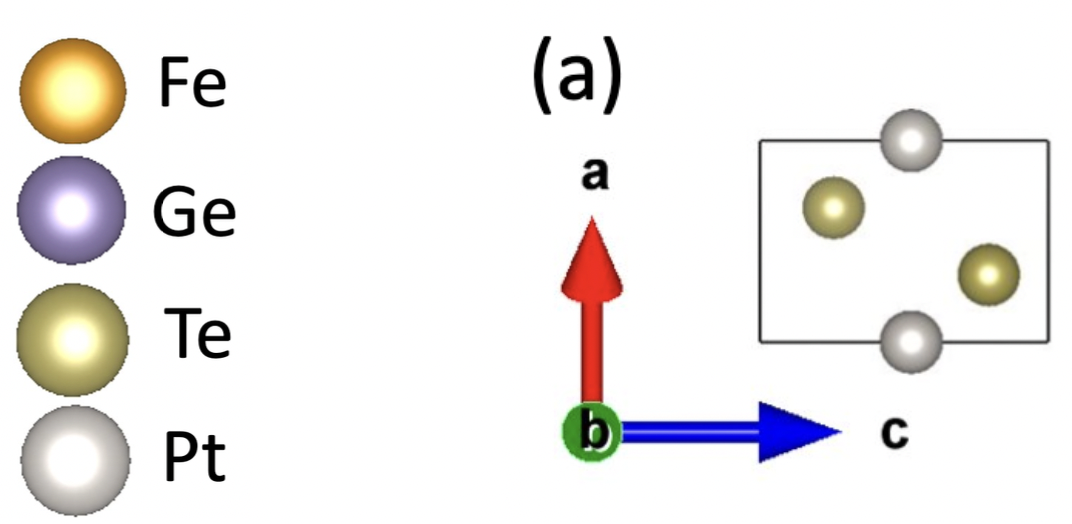}
   \hspace{2.7 cm}
\includegraphics[width=0.17\linewidth]{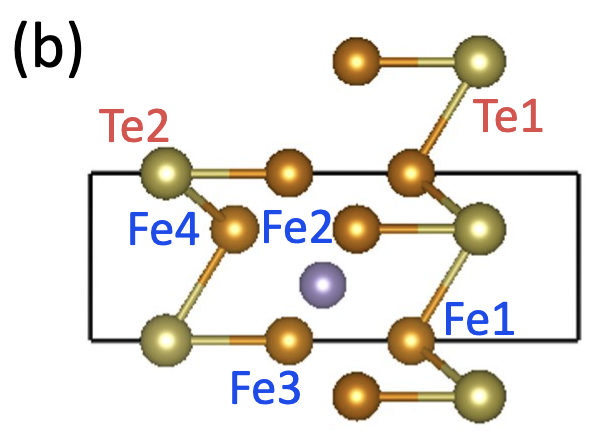}
   \hspace{1.7cm}
\includegraphics[width=0.26\linewidth, trim={0cm -6cm 0cm 0cm}, clip]{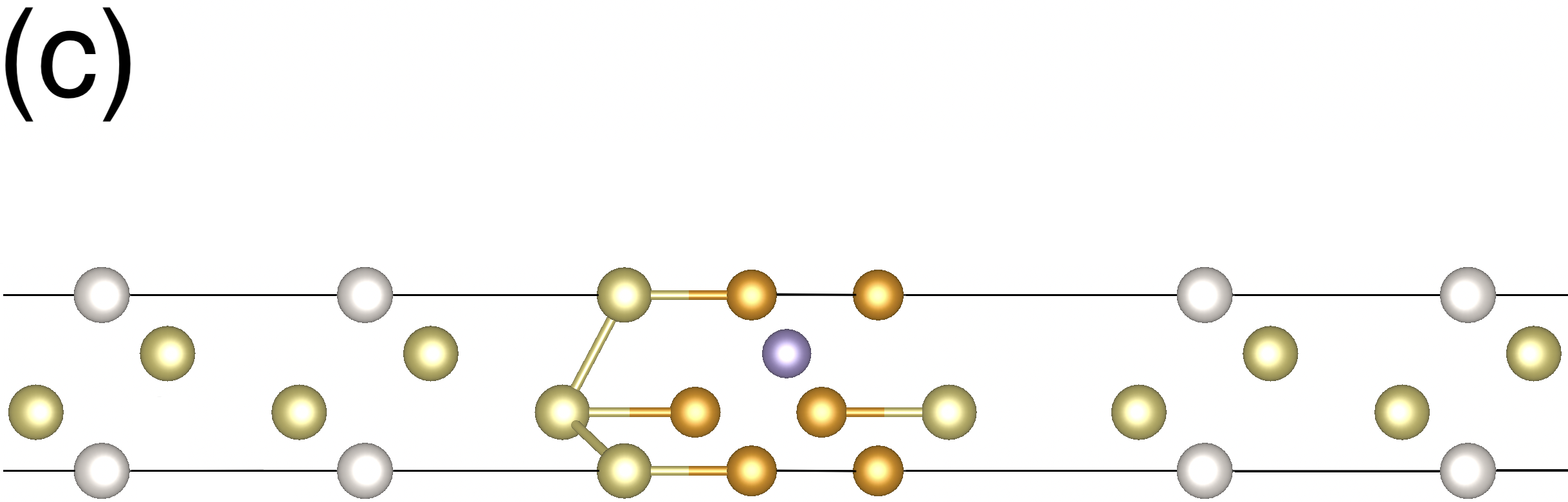}\\
\includegraphics[width=0.08\linewidth, trim={0cm -9.8cm 0cm 0cm}, clip]{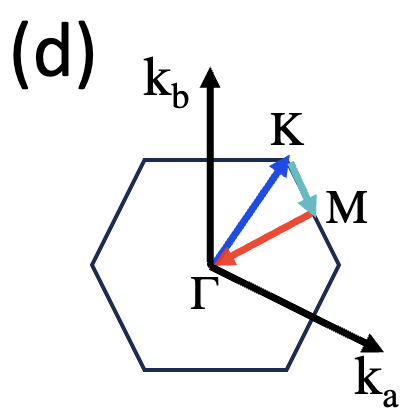}
\includegraphics[width=0.3\linewidth]{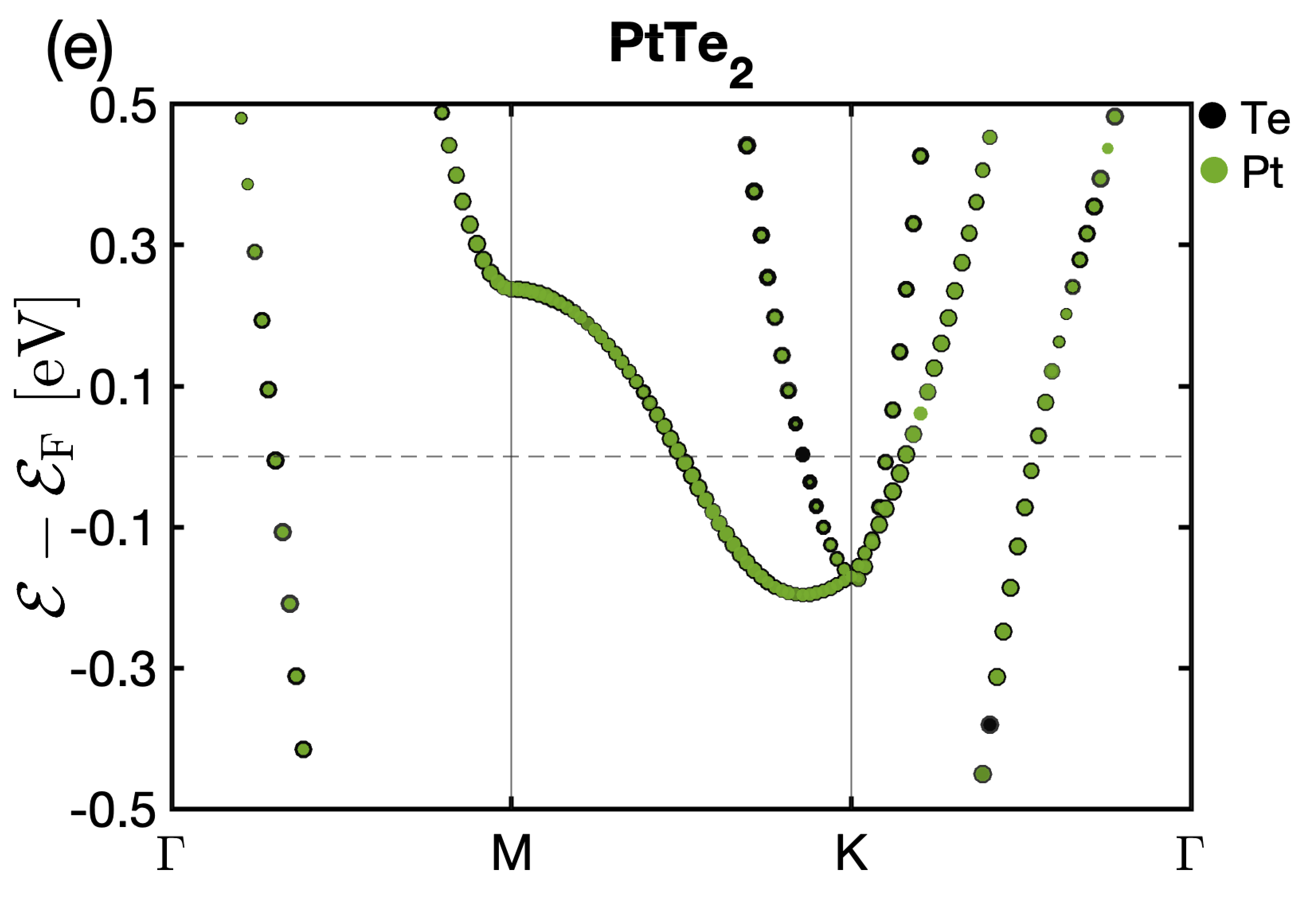}
\includegraphics[width=0.3\linewidth]{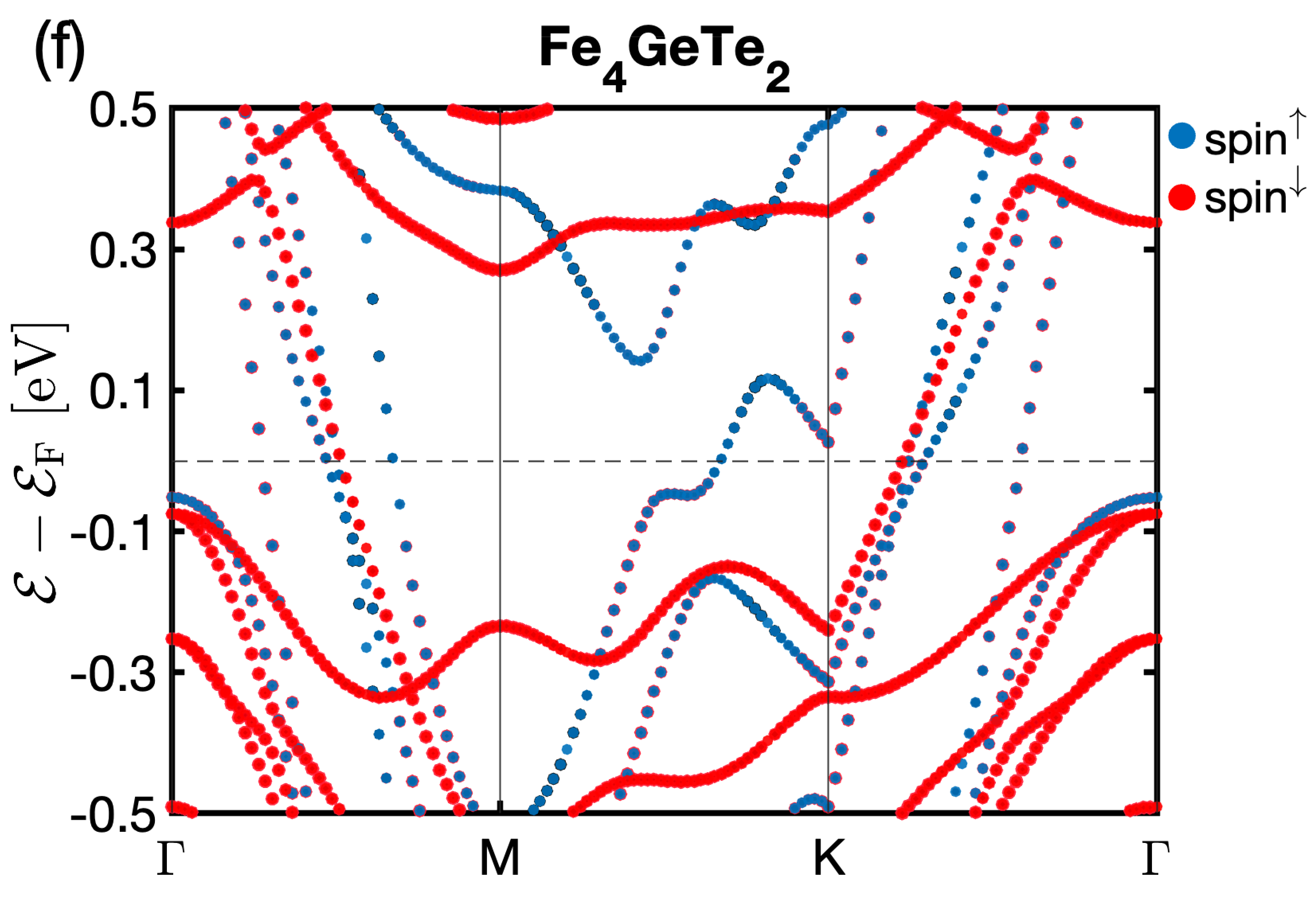}
\includegraphics[width=0.3\linewidth]{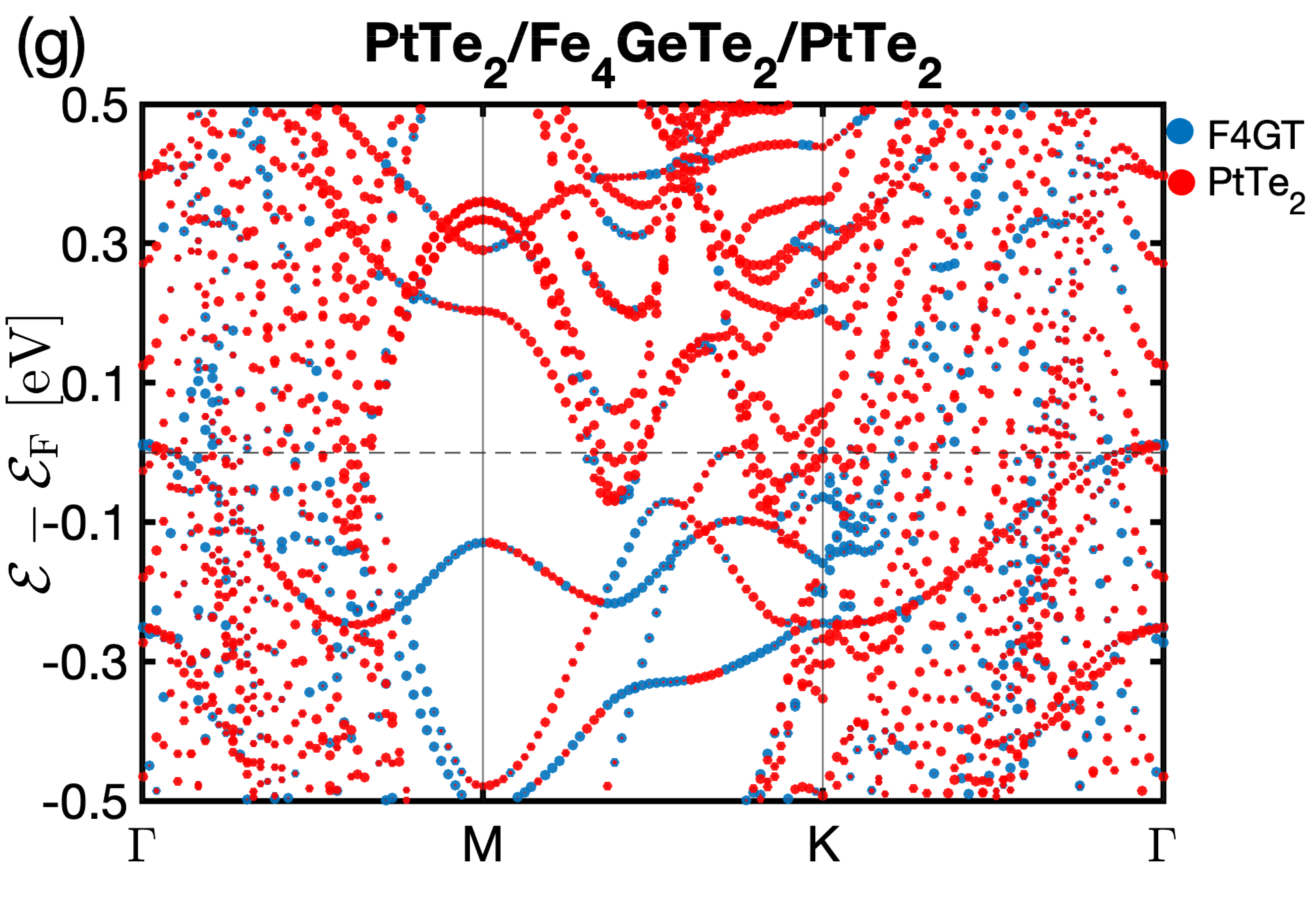}
\caption{\raggedright The atomic structures of (a) PtTe$_2$, (b) F4GT, and (c) F4GT sandwiched between two PtTe$_2$ electrodes. (d) The first Brillouin zone of these structures. (e-g) showcase corresponding electronic band structures.}
    \label{f1}
\end{figure*}

Recent discoveries of Fe$_n$GeTe$_2$ (n=3, 4, 5) (FGT), a class of 2D itinerant ferromagnets with a Curie temperature approaching room temperature, provide exciting prospects for 2D spintronics advancements~\cite{Deiseroth, Zhao, Wang2023,Ershadrad2022-ta}.
A comparative study of the FGT family has been done in Ref. \onlinecite{Ghosh2023} with the aid of ab-initio calculations.
FGT exhibits a notable advantage stemming from its metallic nature, which facilitates the manipulation of both electronic spin and charge. Furthermore, FGT has been suggested as a rare-earth-free material with strong magnetism and electronic correlation~\cite{PhysRevB.93.144404}.
Recent experimental reports have confirmed that the tunneling resistance behavior in hBN sandwiched between two Fe$_3$GeTe$_2$(F3GT) layers follows established patterns, exhibiting minimum (maximum) resistance when the magnetizations of the electrodes are parallel (antiparallel). Remarkably, a significant magnetoresistance of 160$\%$ is observed at low temperatures, indicating a spin polarization of 0.66 in F3GT~\cite{Wang2018}.
Moreover, the formation of Ohmic contacts in $F3GT/MoS2$ interfaces is confirmed by linear current-voltage curves, indicating a conducting layer rather than a tunneling one. 
This is a positive result as Ohmic contacts enable efficient charge transport with minimal resistance, whereas tunneling contacts can obstruct current flow and reduce device performance.
It has been also observed that magnetoresistance of F3GT/MoS$_2$/F3GT heterostructures reaches 3.1$\%$ at 10 K, which is approximately 8 times larger than conventional spin valves with MoS$_2$ and conventional ferromagnetic electrodes~\cite{Lin2020}. While F3GT has been extensively explored, the focus on Fe$_4$GeTe$_2$ (F4GT) has been relatively limited, despite its higher Curie temperature, making it a promising avenue for further research~\cite{doi:10.1126/sciadv.aay8912}.
Investigating F4GT can provide valuable insights into its exceptional electronic, magnetic, and structural properties. By understanding the distinctive characteristics of F4GT, one can unlock its potential for technological applications such as spintronics, magnetic storage, and other advanced electronic devices. 

Here, ab-initio simulations were utilized to analyze the transport characteristics of vdW heterostructures consisting of F4GT. The focus was also on investigating the electronic and magnetic properties of these heterostructures. Specifically, we investigated spin-dependent ballistic transport in both mono- and bi-layer F4GT structures that were sandwiched between PtTe$_2$ electrodes. To assess the tunneling magnetoresistance behavior, we analyzed the spin-dependent electronic transport across $F4GT/GaTe/F4GT$ junctions, connected to PtTe$_2$ electrodes, serving as vdW MTJs.

\section{Methodology}
We employed the QuantumATK software package~\cite{Smidstrup_2020} to investigate quantum transport properties. The calculations involved the combination of density functional theory (DFT) and the nonequilibrium Green's function (NEGF) formalism. The DFT calculations were carried out using the generalized gradient approximation (GGA) to describe the exchange-correlation functional. We obtained the electronic band structure and density of states (DOS) of F4GT to gain insights into its electronic properties. Realistic device structures, including the scattering region and leads, were constructed to simulate the transport properties. The NEGF formalism was employed to calculate the transmission spectra and current-voltage characteristics of the F4GT-based devices under external biases. To accurately account for vdW interactions, we applied the DFT-D3 method with Becke-Jonson damping~\cite{10.1063/1.3382344,https://doi.org/10.1002/jcc.21759}. Structural relaxations were performed using the linear combination of atomic orbitals (LCAO) basis set with PseudoDojo for pseudopotentials. A Monkhorst-Pack grid of 14$\times$14$\times$1 has been used and a cutoff density of 140 Hartree was chosen to ensure convergence. Moreover, a force tolerance of $10^{-3}$ eV/\AA~ was used for relaxations. Convergence was achieved when the total energy difference between consecutive steps was below $10^{-4}$ eV. The source and drain electrodes were set to a length of 10.402 \AA, and the same LCAO settings were applied for the quantum transport simulations. For the gate-all-around (GAA) structure, the Poisson solver utilized the Dirichlet boundary condition in all directions. We utilized a Monkhorst-Pack grid with dimensions of 16$\times$16$\times$1 to assess the transmission and current. To maintain reasonable simulation times, the parallel conjugate gradient method was employed. To calculate the magnetocrystalline anisotropy energy (MAE), we included spin-orbit coupling (SOC) using a full k-point grid and the Brillouin zone integration was performed using a 55$\times$55$\times$1 $\Gamma$-centered Monkhorst-Pack grid.
Moreover, previous studies~\cite{Ghosh2023} have demonstrated that GGA + U calculations are incompatible with experimental results for FGT materials. This incompatibility extends to parameters such as unit cell dimensions, magnetic moments, magnetic anisotropy energy, and transition temperature. Consequently, utilizing static electron correlation is not suitable for accurately characterizing a metallic magnet like FGT. Therefore, we have neglected the Hubbard U correction in our calculations.

The spin-dependent transmission coefficient was determined using Green's functions, as expressed by the following equation~\cite{10.1063/1.5009744}
\begin{equation}
    T_\sigma\equiv Tr\left[Im\left(\Gamma_L^r\right)G^rIm\left(\Gamma_L^r\right)G^a\right],
\end{equation}
The subscript $\sigma\equiv \uparrow\downarrow$ represents the spin index. The term $\Gamma_{\{L,R\}}=\texttt{i}[\Sigma_{\{L,R\}}-\Sigma_{\{L,R\}}^{\dagger}]$ is the line width function and $\Sigma_{L(R)}^r$ in the equation corresponds to the retarded self-energy of the left (right) electrode, representing the coupling between the central region and the semi-infinite leads. This term accounts for the interaction and exchange of electrons between the central region and the electrodes. The $G^{r(a)}$ term refers to the retarded (advanced) Green's function matrices, which describe the propagation of electrons through the system in the spin and orbital spaces. 

To find the transmission eigenstates, we utilize a linear combination of Bloch states, $\sum_n e_{\alpha n} \psi_{n}$, with coefficients $e_{\alpha n}$ that diagonalize the transmission matrix. This transmission matrix is mathematically defined as
\begin{equation}
    T_{m,n}=\sum_k t_{nk} t^\dagger_{km}
\end{equation}
Here, $t_{nk}$ represents the transmission amplitude, indicating how likely an electron in Bloch state $\psi_{n}$ on the left electrode will have a transition to Bloch state $\psi_{k}$ on the right electrode. The transmission coefficient can be computed by taking the trace of $T_{m,n}$.
 
To calculate the spin-dependent tunneling current, the Landauer-Büttiker formula was employed~\cite{datta_2005},
\begin{equation}
    I_\sigma(V)=\frac{2e}{h} \int d\mathcal{E}\, T_\sigma(\mathcal{E},V_L,V_R)[f_L(\mathcal{E},\mu_L)-f_R(\mathcal{E},\mu_R)],\,
    \label{eqLB}
\end{equation}
 The tunneling current is determined by the electrochemical potentials $\mu_L(\mu_R)$, Fermi distribution functions $f_L(f_R)$, and bias voltages $V_L(V_R)$ applied to the left and right leads at room temperature. The transmission coefficient $T_\sigma(\mathcal{E}, V_L, V_R)$ is energy-dependent and varies with the bias voltages and energy of the system. We performed the current calculations at room temperature (300 K).

\section{Results and Discussion}

\subsection{Characterization of PtTe$_2$/F4GT/PtTe$_2$ heterostructure}
The unit cell of PtTe$_2$ is depicted in Figure~\ref{f1}a. It possesses a layered crystal structure within the trigonal space group P-3m1~\cite{Casado_Aguilar_2022}. The structure comprises Pt atoms situated between two layers of Te atoms. The stacking of these layers repeats in a hexagonal pattern along the c-axis, generating a three-dimensional structure. The interlayer interactions are governed by weak vdW forces. The lattice constants obtained using the GGA functional are a=b=4.01~\AA~ and c=5.201~\AA~. Figure~\ref{f1}b~indicates the layered crystal structure of F4GT, which shares the same space group (P-3m1) as PtTe$_2$~\cite{Liu2022}. Each layer consists of four Fe atoms surrounded by Ge and Te atoms. The Ge and Te atoms form a distorted hexagonal lattice, with the Fe atoms occupying the centers of distorted octahedra formed by the coordination with Ge and Te.
Magnetic moment of Fe1~(Fe4) and Fe2~(Fe3) have been calculated to be 2.72 and 1.71 $\mu_{B}$, respectively, which is in good agreement with Ref.~\onlinecite{Kim2021}.
The calculated lattice constant of F4GT is found to be a=b=3.968~\AA~, which is remarkably close to the lattice constant of PtTe$_2$. This similarity in lattice constants suggests a strong structural resemblance between F4GT and PtTe$_2$, indicating potential similarities in their crystal structures and bonding arrangements. We initiate our investigation by examining a two-probe system consisting of a single-layer F4GT situated between two PtTe$_2$ electrodes (Figure~\ref{f1}c). In this setup, the F4GT layer is subjected to a tensile strain of 0.7$\%$. The distance between the Te atoms on the surface of F4GT and the surface layer of PtTe$_2$ is approximately 2.88~\AA. This distance is larger than the interlayer distances in bulk PtTe$_2$, which is around 2.33~\AA, and smaller than the interlayer distances in bulk F4GT, which is approximately 3.28~\AA. Moreover, 
in the device configuration, the magnetic moment of Fe1 and Fe4 remains unchanged at 2.72 $\mu_{B}$. However, the magnetic moment of Fe2 and Fe3 slightly increases to 1.81 $\mu_{B}$.
\begin{figure}[t]
\centering
\includegraphics[width=.8\linewidth]{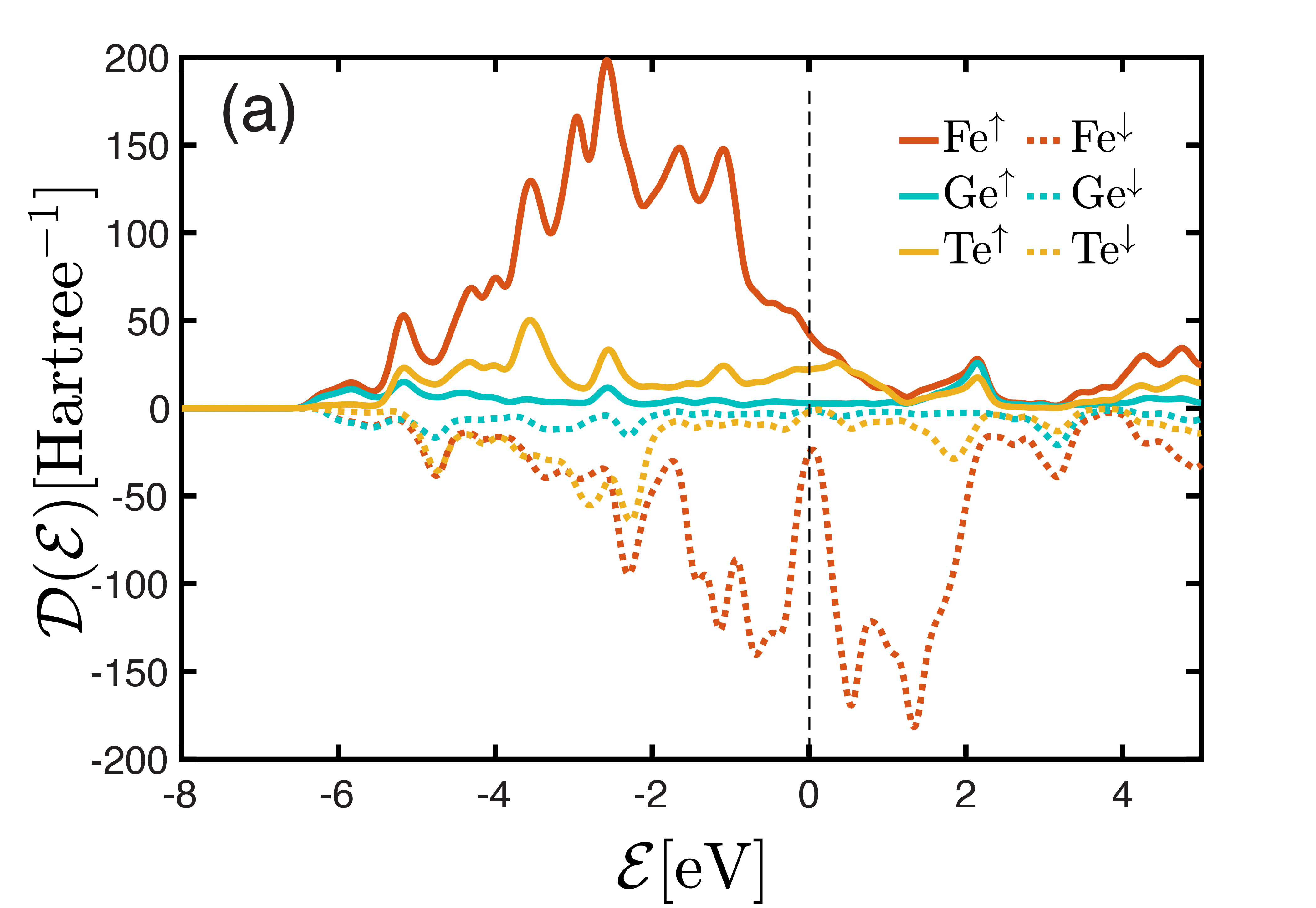}\\
\includegraphics[width=.8\linewidth]{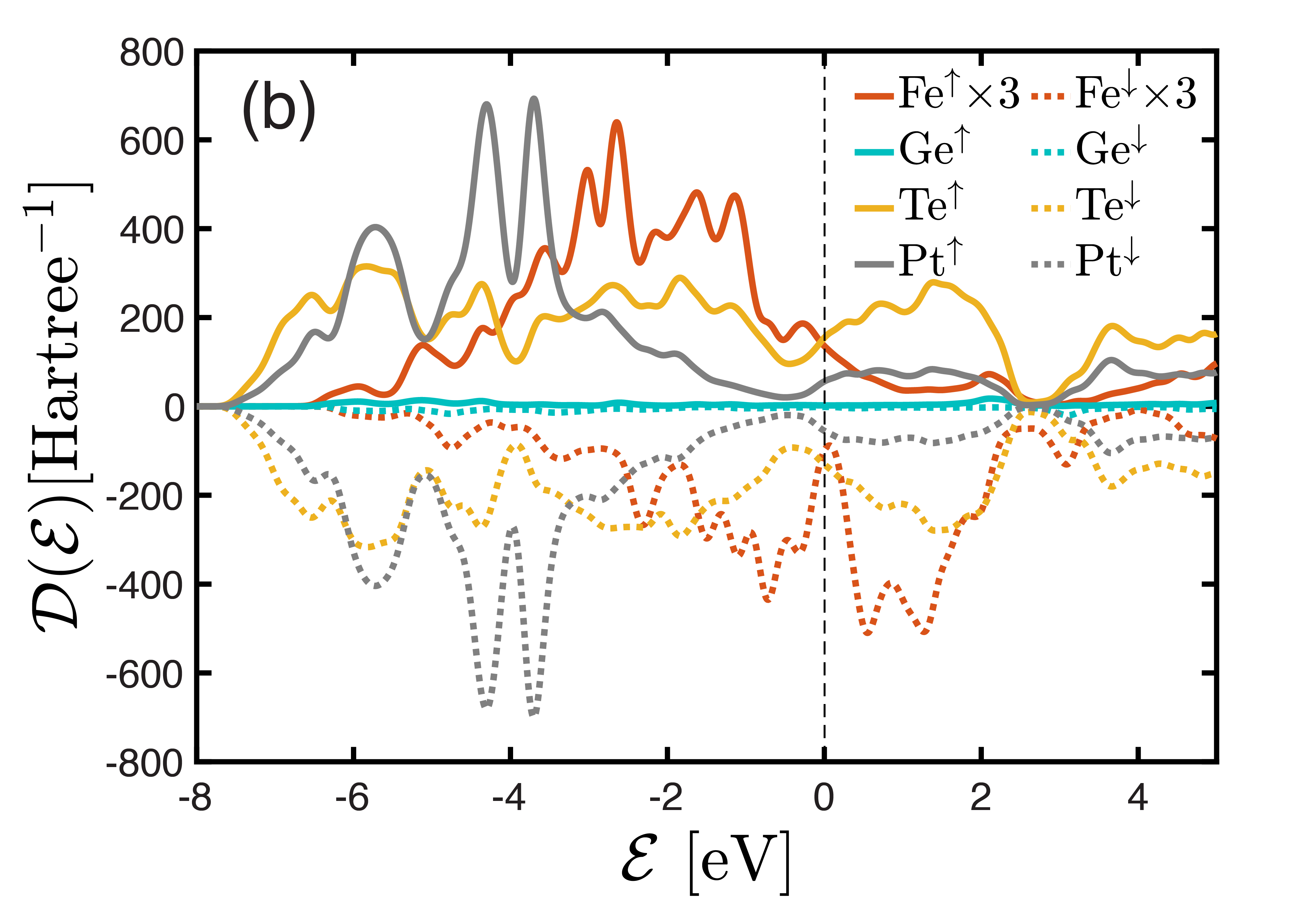}
\caption{\raggedright DOS plots for (a) the freestanding monolayer F4GT and (b) the F4GT sandwiched between two layers of PtTe$_2$.}
    \label{fDOS}
\end{figure}

Figure~\ref{f1}(e) displays the atom-projected band structure plots of the monolayer PtTe$_2$. Additionally, Figure~\ref{f1}(f) presents the spin-polarized bands of F4GT, while Figure~\ref{f1}(g) shows bands projected onto both the total F4GT layer and the PtTe$_2$ layer of the PtTe$_2$/F4GT/PtTe$_2$ system.
The band structure analysis reveals that both monolayer PtTe$_2$ and F4GT exhibit metallic behavior. In the case of F4GT, the states around the Fermi level are primarily dominated by spin-up channel, accompanied by a smaller contribution from spin-down channel. As observed in panel (c) of the weight-projected band structure plot, numerous bands intersect and cross the Fermi level. These crossing points indicate the presence of potential conductance channels within the device configuration. The majority of states near the Fermi level originate from Te and Pt atoms from PtTe$_2$ layers. This difference can be attributed to the larger number of Te and Pt atoms present in the electrodes of the device, leading to a higher contribution from these elements to the electronic states near the Fermi level. Furthermore, upon closer examination near the Fermi level at the M and K points (panel c), it is evident that the F4GT bands in the device configuration exhibit minimal changes compared to the isolated F4GT (panel b). This observation suggests a weak interaction at the interface between F4GT and the electrodes. The relatively unchanged Fe bands indicate that the electronic structure of F4GT is preserved within the device, implying that the interface interaction does not significantly affect the Fe electronic states.

The spin-polarized DOS calculations were performed to investigate the spin-dependent electronic properties of the monolayer F4GT (a) and the F4GT sandwiched between two PtTe$_2$ electrodes (b), as shown in Fig.~\ref{fDOS}. Panel (a) reveals a significant electron density at the Fermi level ($\mathcal{E}_F$) with an exchange splitting, indicating the ferromagnetic nature of F4GT. This finding is consistent with previous ab initio calculations~\cite{Kim2018}. The states near the Fermi level are primarily dominated by the Fe atoms in F4GT, contributing to the majority spin channel. In panel~(b), the DOS plot showcases the impact of the PtTe$_2$ layers on the electronic states of the F4GT sandwiched structure. It is observed that the redistribution and shifting of the electronic states in the Fe atoms are relatively unchanged, indicating a weak interaction between the F4GT and PtTe$_2$ layers. This suggests that the electronic properties of F4GT remain largely preserved within the device configuration. Furthermore, the PDOS values from panel (b) highlight the spin filtering mechanism of the device configuration. The electronic states at $\mathcal{E}_F$ in the majority spin channel are more abundant compared to the minority spin channel. This indicates the potential for spin-polarized currents when a bias voltage is applied to the device.

\begin{table}[b]
    \caption{\raggedright Comparison of the atom projected MAE (in unit of mJ/m$^2$) for mono- and bi-layer freestanding F4GT with the device configuration, where F4GT is sandwiched between two layers of PtTe$_2$. Values in parentheses for bilayer cases refer to the MAE of corresponding atoms in the second layer.}
    \centering
    \begin{tabular}{ccccc}
    \hline
\multicolumn{1}{c}{} & \multicolumn{2}{c}{\textbf{Monolayer}} & \multicolumn{2}{c}{\textbf{Bilayer}} \\
\cmidrule(rl){2-3} \cmidrule(rl){4-5}
\textbf{ } & {freestanding    } & {device\,\,\,\,\,\,\,\,\,\,\,\,} & {freestanding  } & {device}  \\
\hline
 Fe1 & -1.08 & -0.78 & -0.67 (-0.9) &  -0.62 (-0.857)\\
 Fe2&-0.44  &-0.33  & -0.40 (-0.39)  & -0.41 (-0.34) \\
 Fe3&-0.44  & -0.33 & -0.39 (-0.41) &-0.37 (-0.39)  \\
 Fe4&-1.12  &-0.78  & -0.89 (-0.68) &  -0.93 (-0.57)\\
 Te1&-0.35 &-0.12 & -0.24 (-0.45) &-0.02 (-0.29)\\
 Te2&-0.38&-0.12& -0.45 (-0.25)&-0.36 (-0.03)\\
 Ge&-0.03& -0.03&  0.005 (0.004)&0.02 (0.02)\\
 \hline
 \textbf{Total} & \textbf{-3.84} & \textbf{-2.17} &\textbf{-6.09}&  \textbf{-4.55}\\
\hline
    \end{tabular}
    \label{tmae}
\end{table}

\begin{figure}[t]
\centering
\includegraphics[width=1\linewidth]{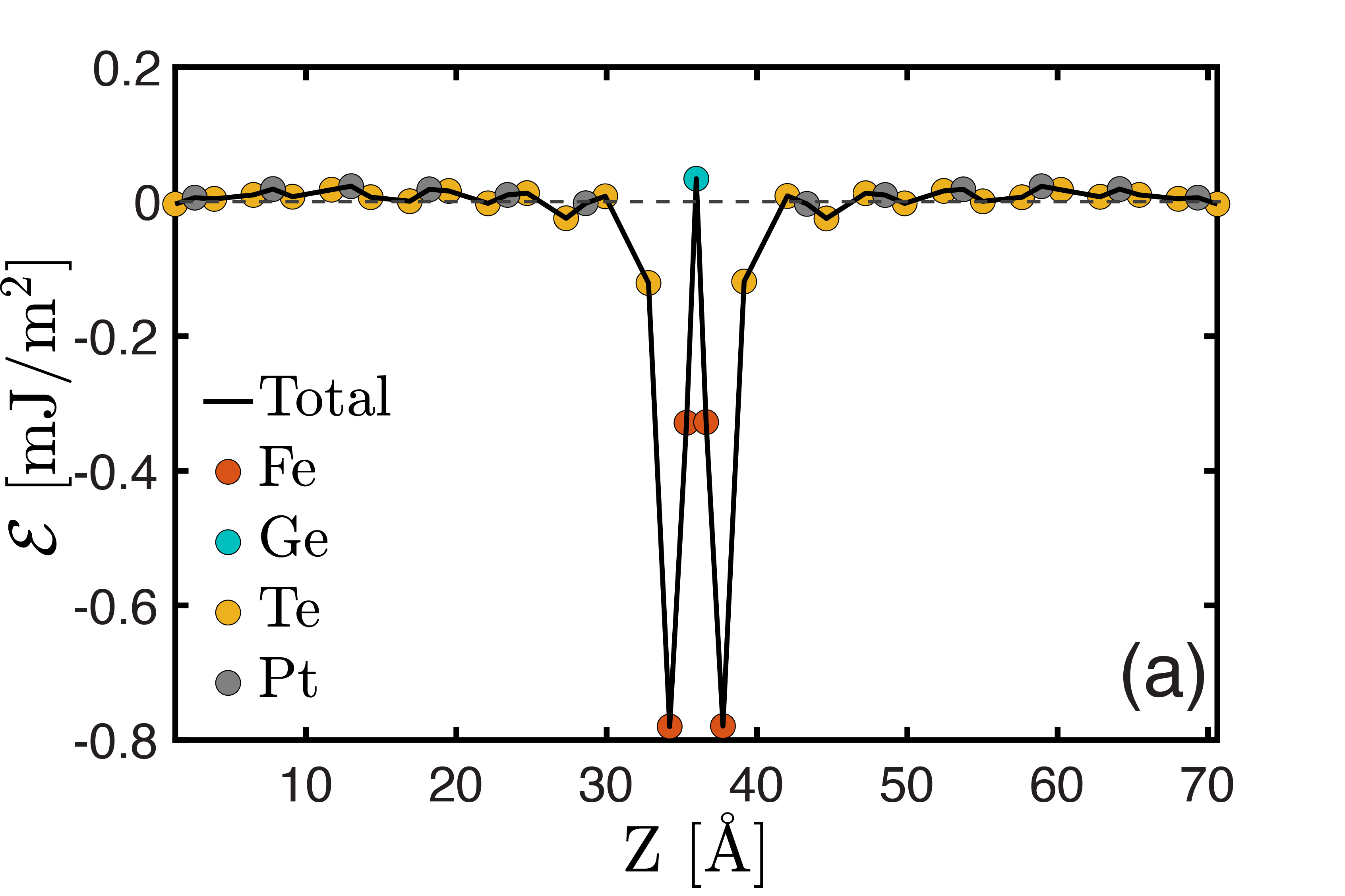}\\
\includegraphics[width=1\linewidth]{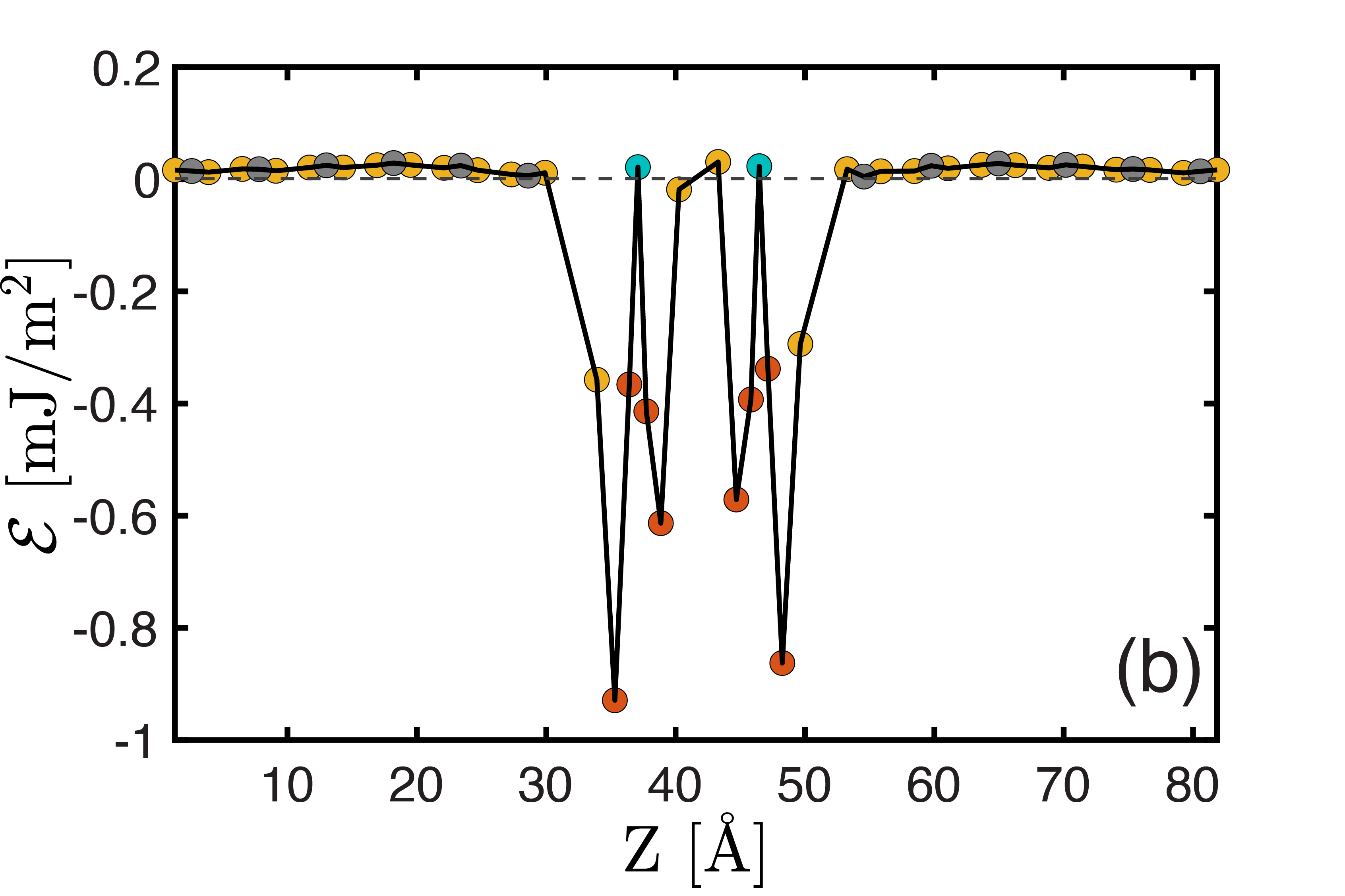}
\caption{\raggedright Atom projected in-plane magnetocrystalline anisotropy energy plots for (a) mono- and (b) bi-layer F4GT sandwiched between two PtTe$_2$ electrodes.}
    \label{fMAE}
\end{figure}

In spintronic devices, the behavior of charge transport is highly dependent on how the magnetic moments in the materials are aligned. To investigate this relationship, we calculated the MAE of (a)~single- and (b)~bi-layer F4GT connected to PtTe$_2$ electrodes in Fig.~\ref{fMAE}. The MAE values were calculated using the force theorem~\cite{PhysRevB.41.11919}. The MAE quantifies the energy difference between two spin orientations aligned along the easy axis (the preferred direction) and the hard axis (the unfavorable direction) of the material, which are described by the spherical angles $\theta$ and $\phi$, $MAE=\mathcal{E}(\theta_1,\phi_1)-\mathcal{E}(\theta_0,\phi_0)$]. 
The obtained results demonstrate a substantial in-plane MAE at the scattering part for the device configurations. This finding aligns with the reported MAE value of freestanding F4GT, which also indicates an in-plane easy axis in various calculations~\cite{Kim2022, PhysRevB.104.104427}.
The weak interaction at the interface between F4GT and the electrodes does not change the direction of the MAE.  

For a comprehensive investigation of the electrode's impact on the MAE of F4GT, we calculated the atomic MAE values for both mono- and bi-layer freestanding F4GT, as well as the device configuration with F4GT placed between two layers of PtTe$_2$ (see Tab. \ref{tmae}). All values are given in units of mJ/m$^2$, and for the bi-layer cases, the values in parentheses indicate the MAE of corresponding atoms in the second layer.
We observed that the direction of MAE remains consistent across all cases, indicating that the preferred magnetic orientations in F4GT persist regardless of the presence of electrodes. However, the absolute value of MAE decreases when F4GT is connected to the electrodes. 
This decrease signals the influence of the weak interaction on the stability and alignment of the magnetic moments in F4GT when placed between the electrodes.
A reduction in the MAE of F4GT in the device configuration compared to its freestanding form suggests that the magnetic moments in the material become less stable and more susceptible to changes in external conditions. The lower MAE implies that the magnetic moments in F4GT require less energy to switch their orientation.

To assess the impact of SOC on the transport properties of the device, we have calculated the transmission coefficient for the device composed of F4GT sandwiched between two PtTe$_2$ electrodes. Our findings indicate that SOC has a negligible effect on the transmission coefficient of the device. This can be primarily attributed to the use of non-magnetic electrodes in our calculations. The magnetic properties influencing transport are derived from the Fe atoms in the F4GT layer. For the purpose of focusing on transport phenomena, we intentionally excluded SOC from our consideration in the model.

\subsection{Scattering spin filter}

\begin{figure}[b]
\centering
\includegraphics[width=0.9\linewidth, trim={0cm 0cm 0cm 3cm}, clip]{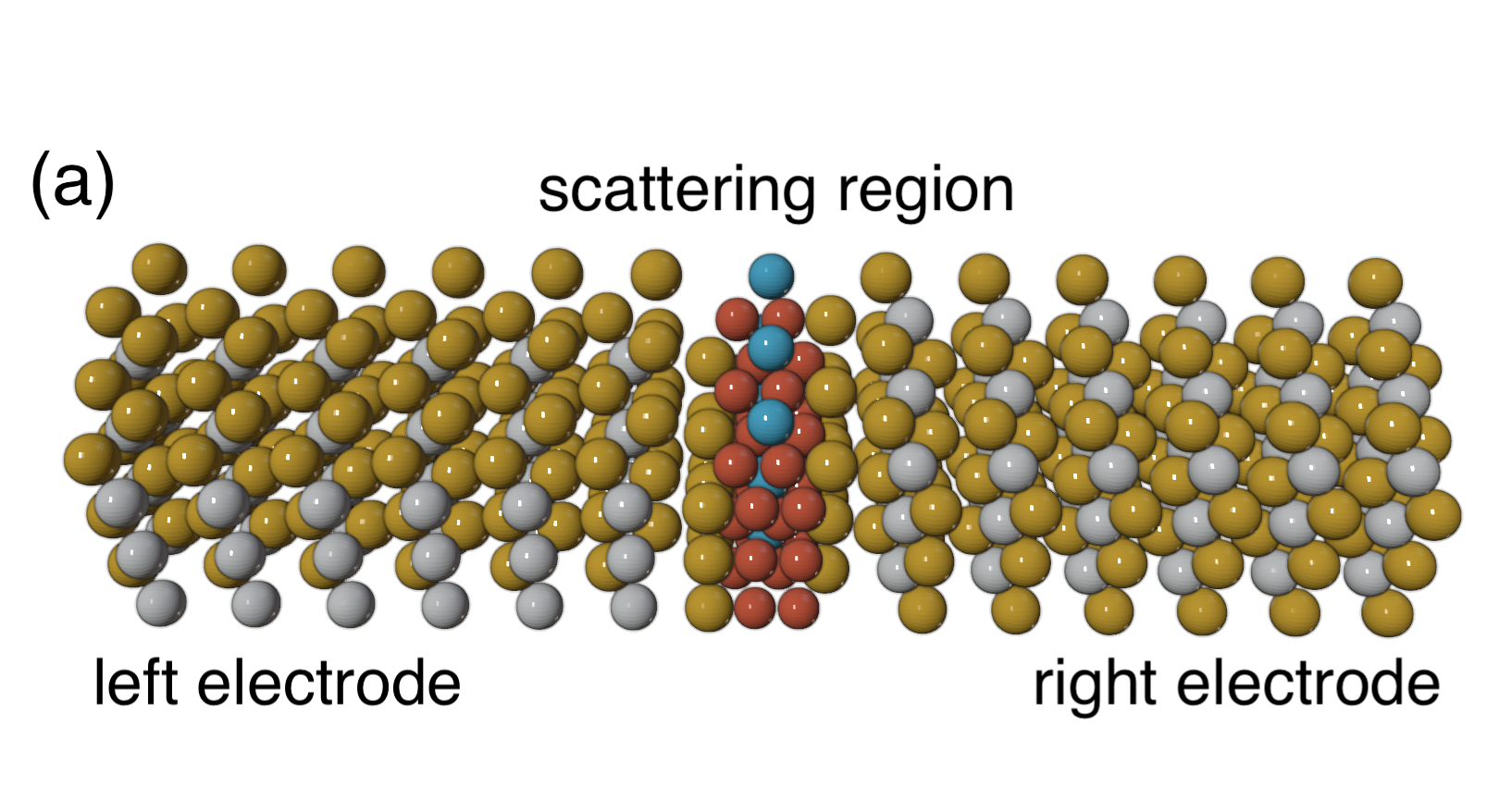}\\
\includegraphics[width=0.49\linewidth]{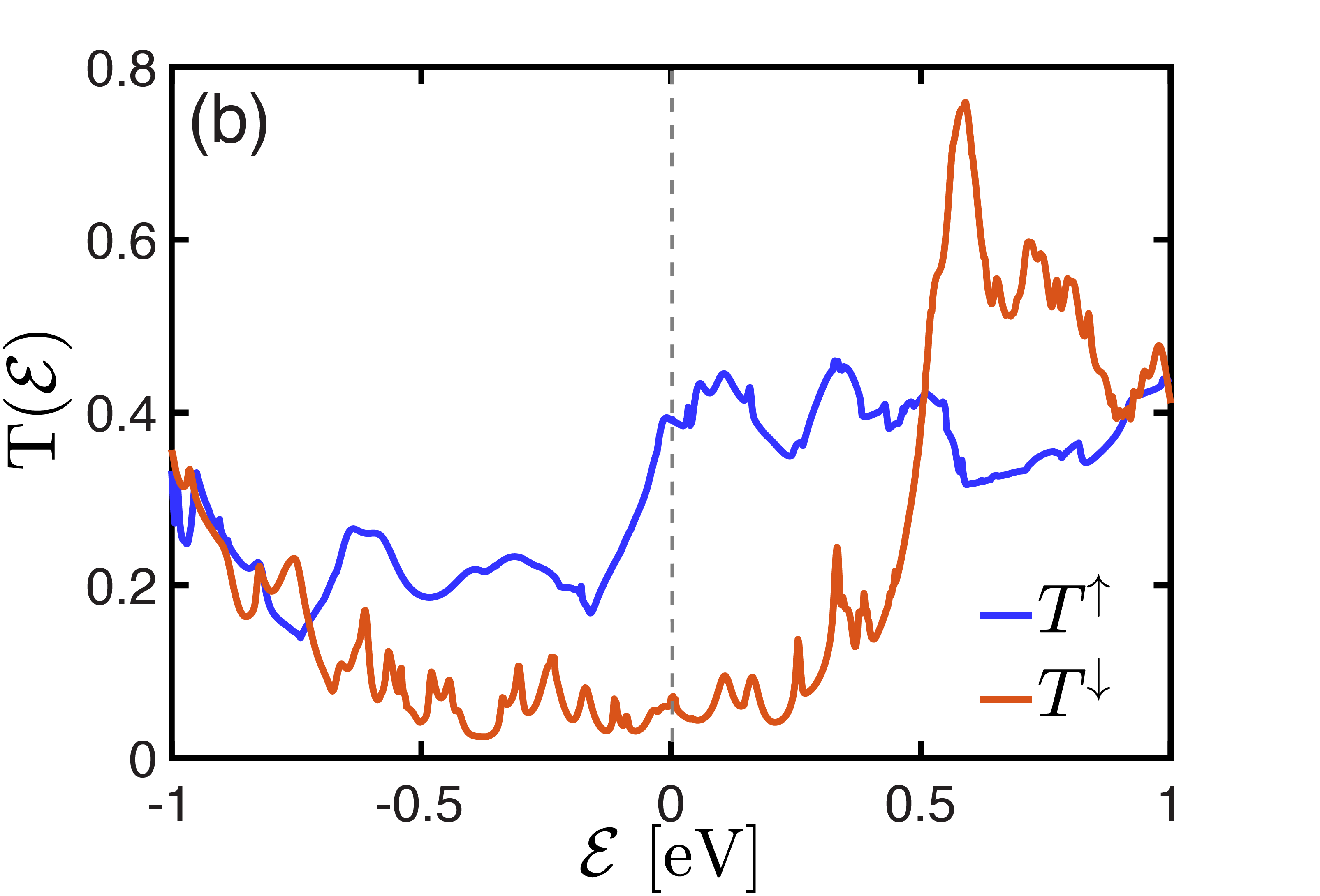}
\includegraphics[width=0.49\linewidth]{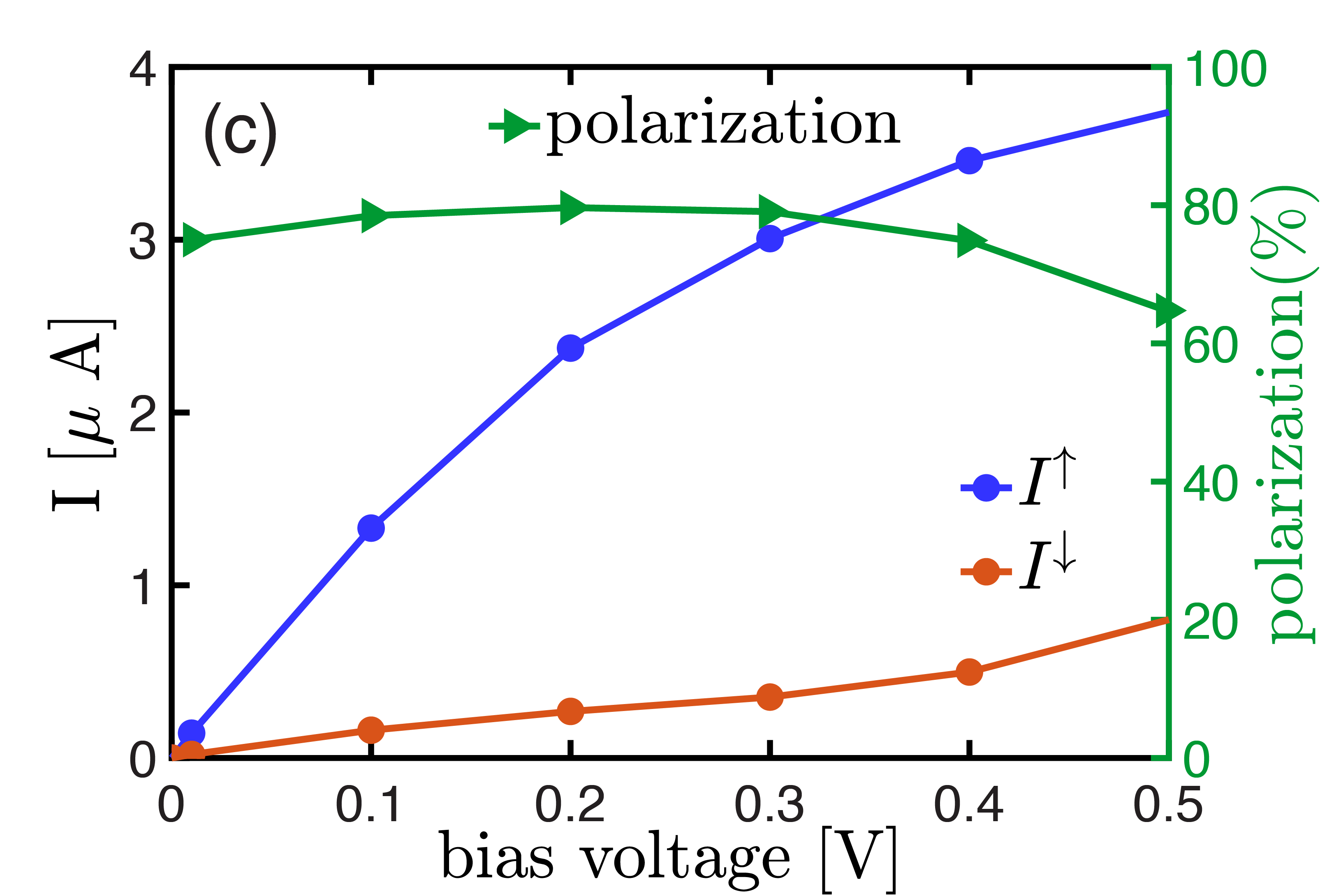}\\
\includegraphics[width=0.49\linewidth]{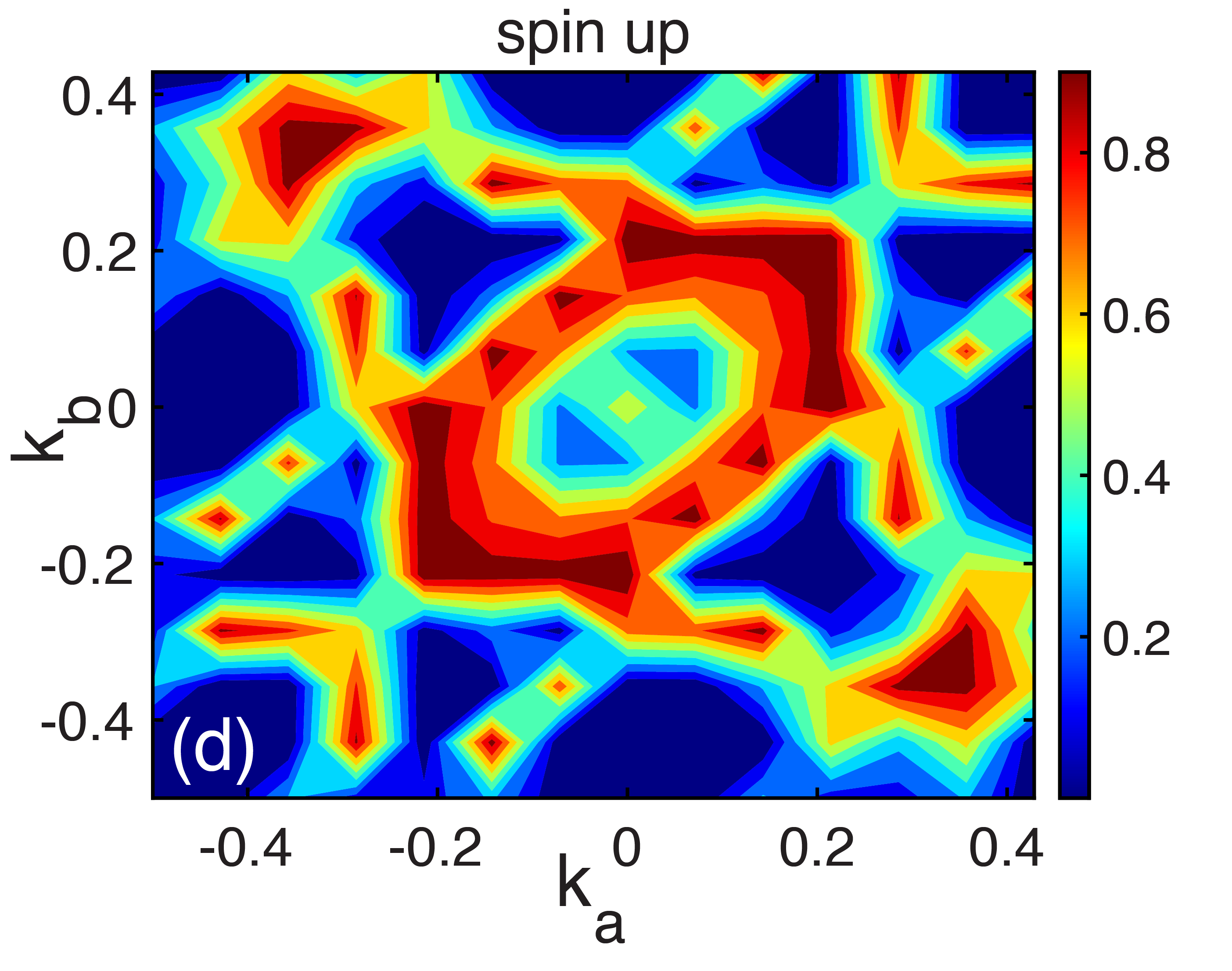}\includegraphics[width=0.49\linewidth]{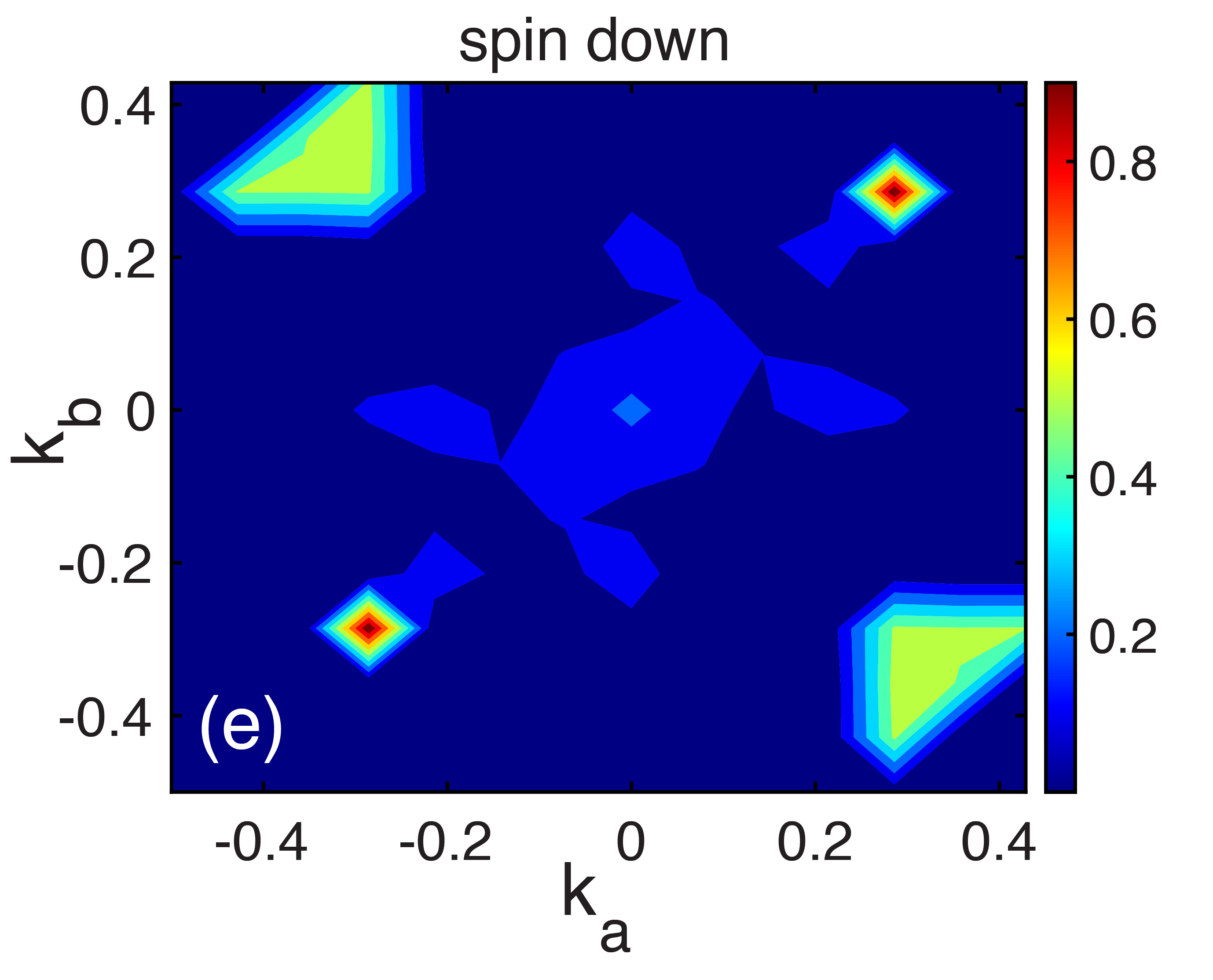}
\caption{\raggedright (a) Schematic representation of a two-probe model for NEGF calculations, depicting a monolayer F4GT sandwiched between two PtTe$_2$ electrodes. (b) The transmission spectrum of the system under zero bias voltage. (c) Variation of spin polarization, spin-up current $(I{\uparrow})$ and spin-down current $(I{\downarrow})$ with bias voltage. The $k_\parallel$-resolved transmission probability for (d) spin-up and (e) spin-down electrons at the Fermi energy in the absence of bias voltage. The path of $k_a$ and $k_b$ in first Brillouin zone is shown in Fig.\ref{f1}d.}
    \label{f_TEmono}
\end{figure}

\begin{figure}[b]
    \centering
    \includegraphics[width=\linewidth, trim={0cm 8cm 0cm 0.2cm}, clip]{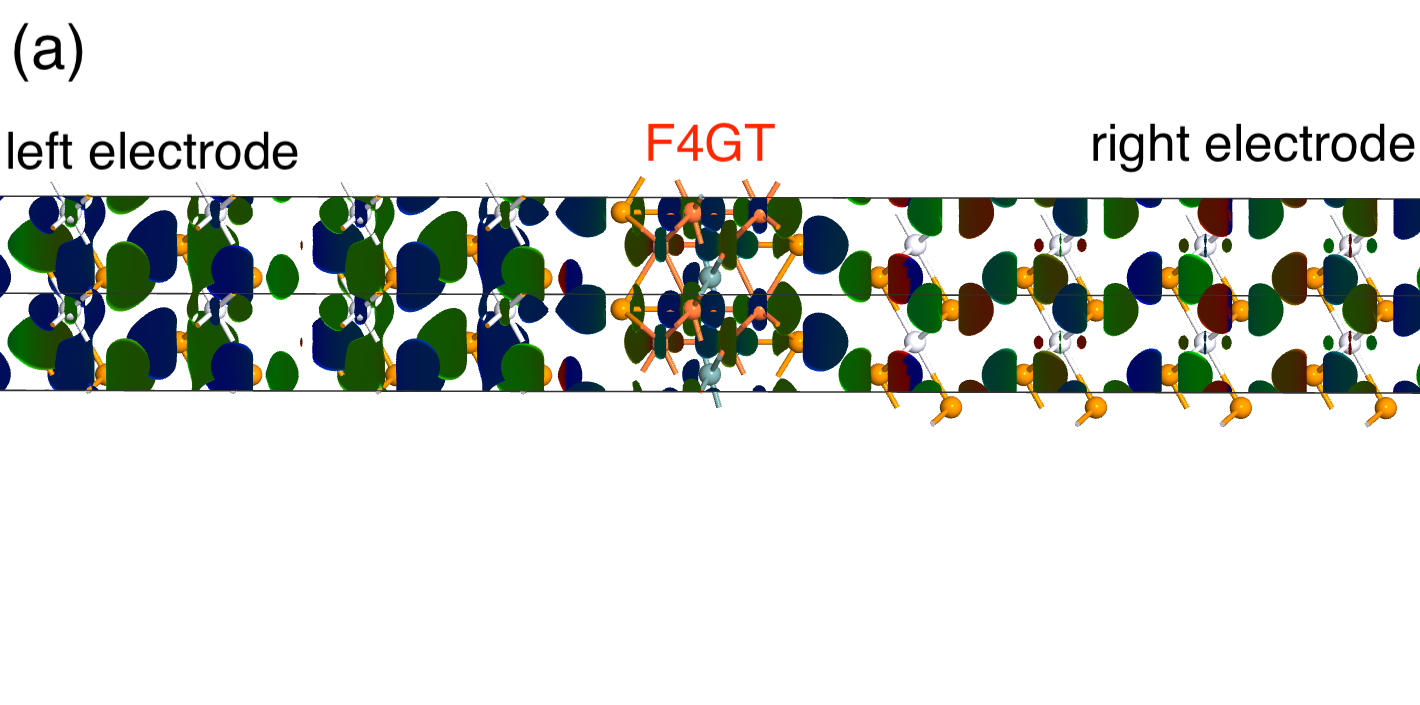}\\
    \includegraphics[width=\linewidth, trim={0cm 4cm 0cm 5cm}, clip]{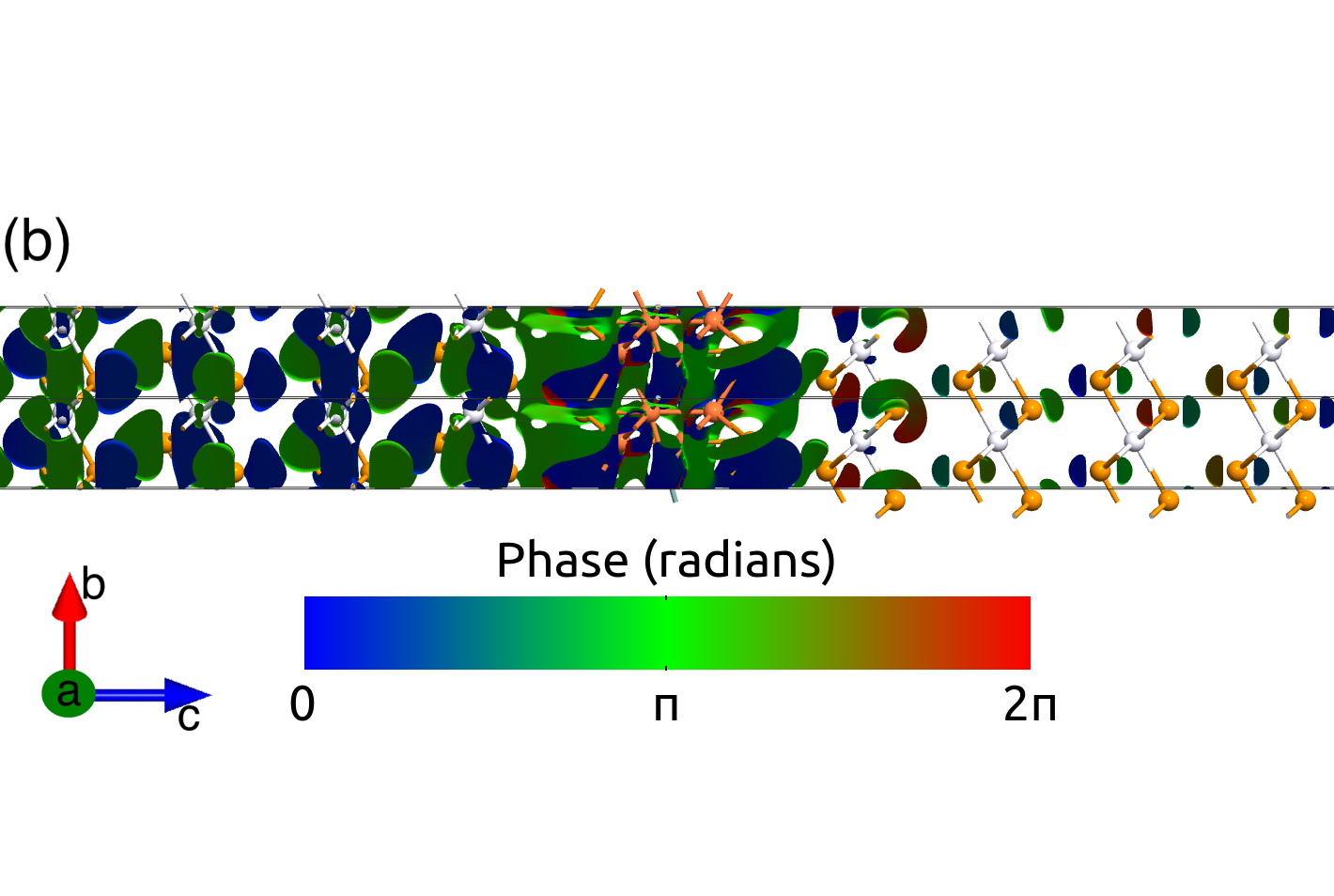}
    \caption{\raggedright The isosurface plots of transmission eigenstates for a monolayer F4GT placed between two PtTe$_2$ electrodes under zero bias voltage at the Fermi level for (a) the spin-up and (b) the spin-down channel. For both channels the isovalues are fixed at 0.17\AA$^{-3}$eV$^{-1}$.}
    \label{eigenTE}
\end{figure}

\begin{figure}
    \centering
    \includegraphics[width=\linewidth]{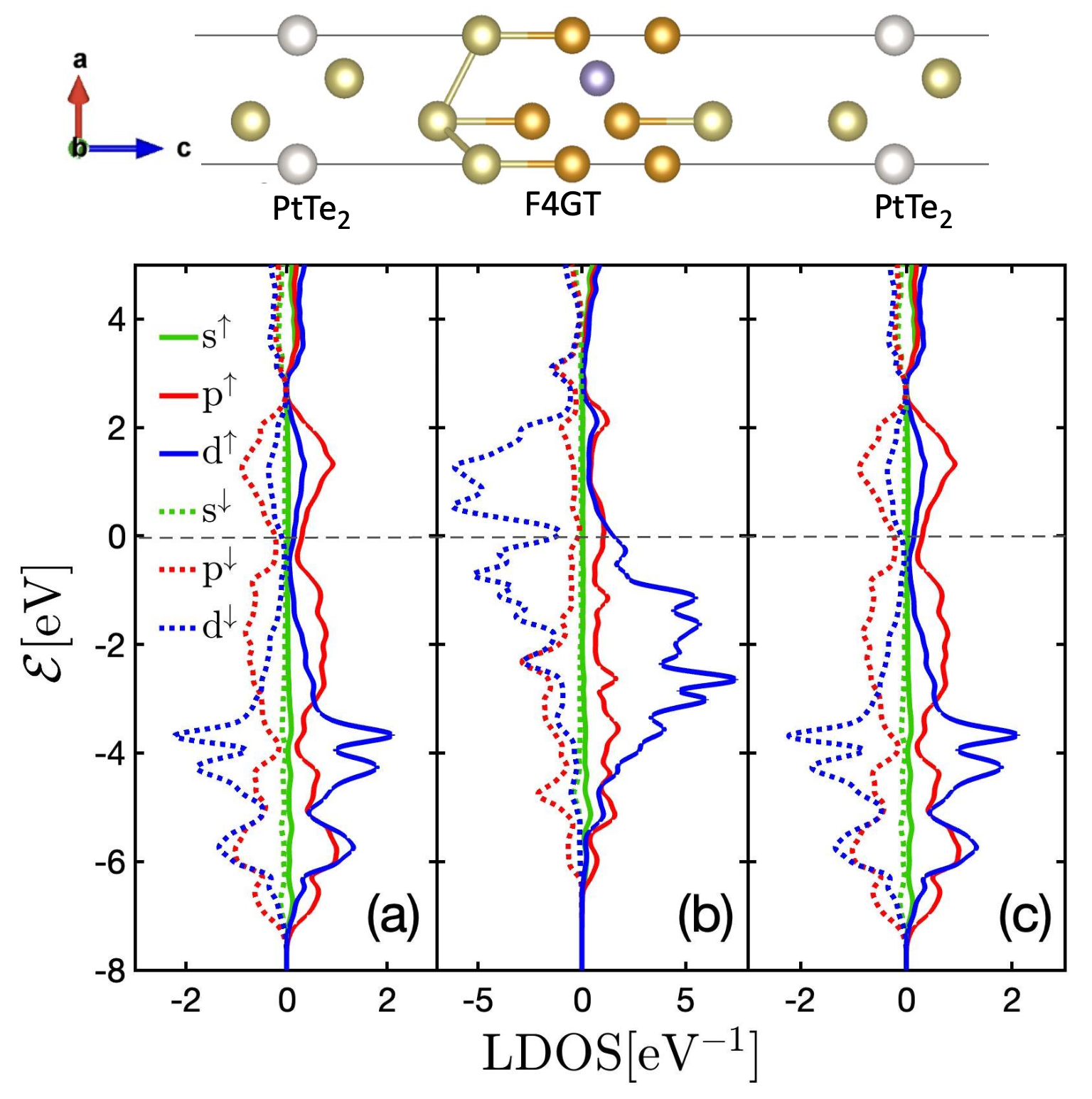}
    \caption{Orbital-projected local density of states for (a) the left electrode's PtTe$_2$ layer interfacing with F4GT, (b) F4GT itself, and (c) the right electrode's PtTe$_2$ layer interfacing with F4GT.}
    \label{ldos}
\end{figure}

\begin{figure*}[t]
\centering
\includegraphics[width=.65\linewidth, trim={0cm -1cm 0cm 2cm}, clip]{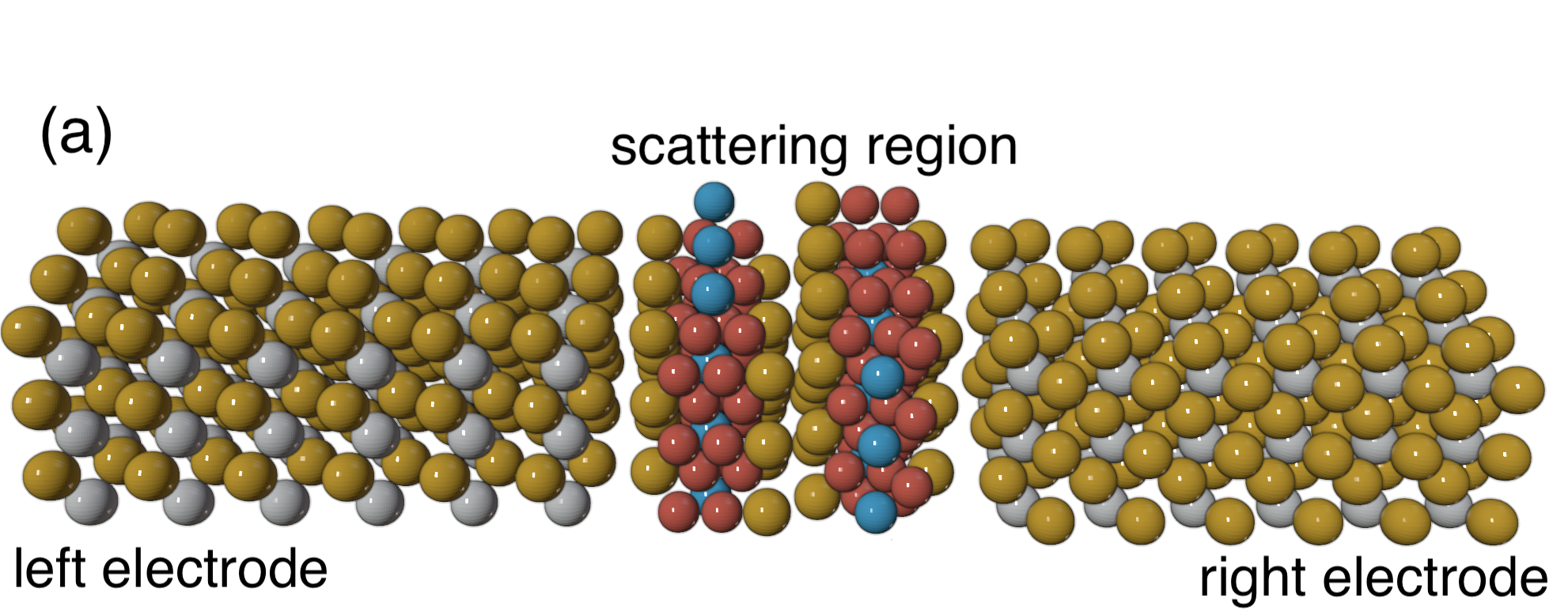}\\
{\large\raggedleft{\textbf{FM configuration}}\hspace{5cm}\raggedleft{\textbf{AFM configuration}}} \\
\includegraphics[width=0.23\linewidth]{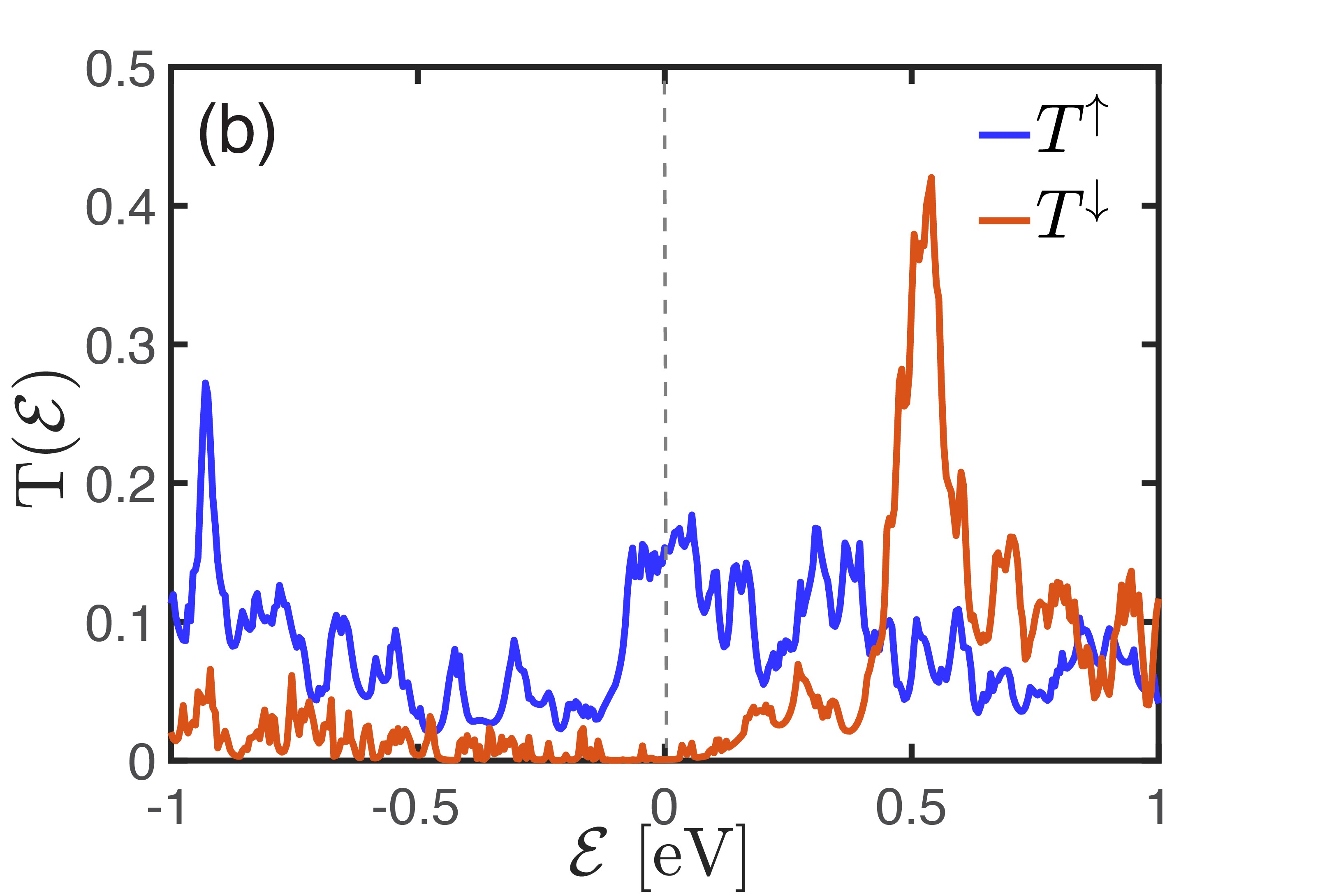}
\includegraphics[width=0.23\linewidth]{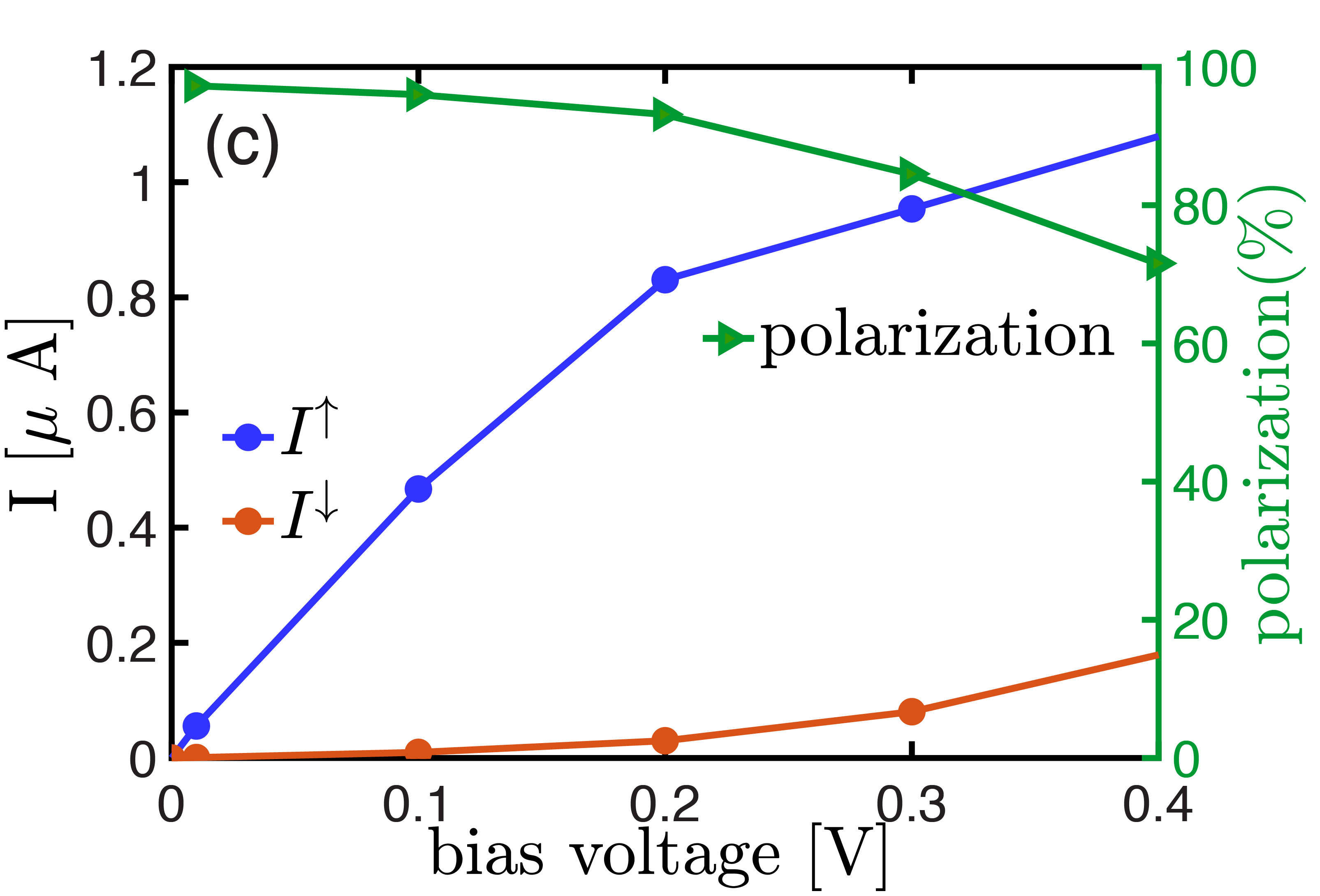}\hspace{0.8cm}\includegraphics[width=0.23\linewidth]{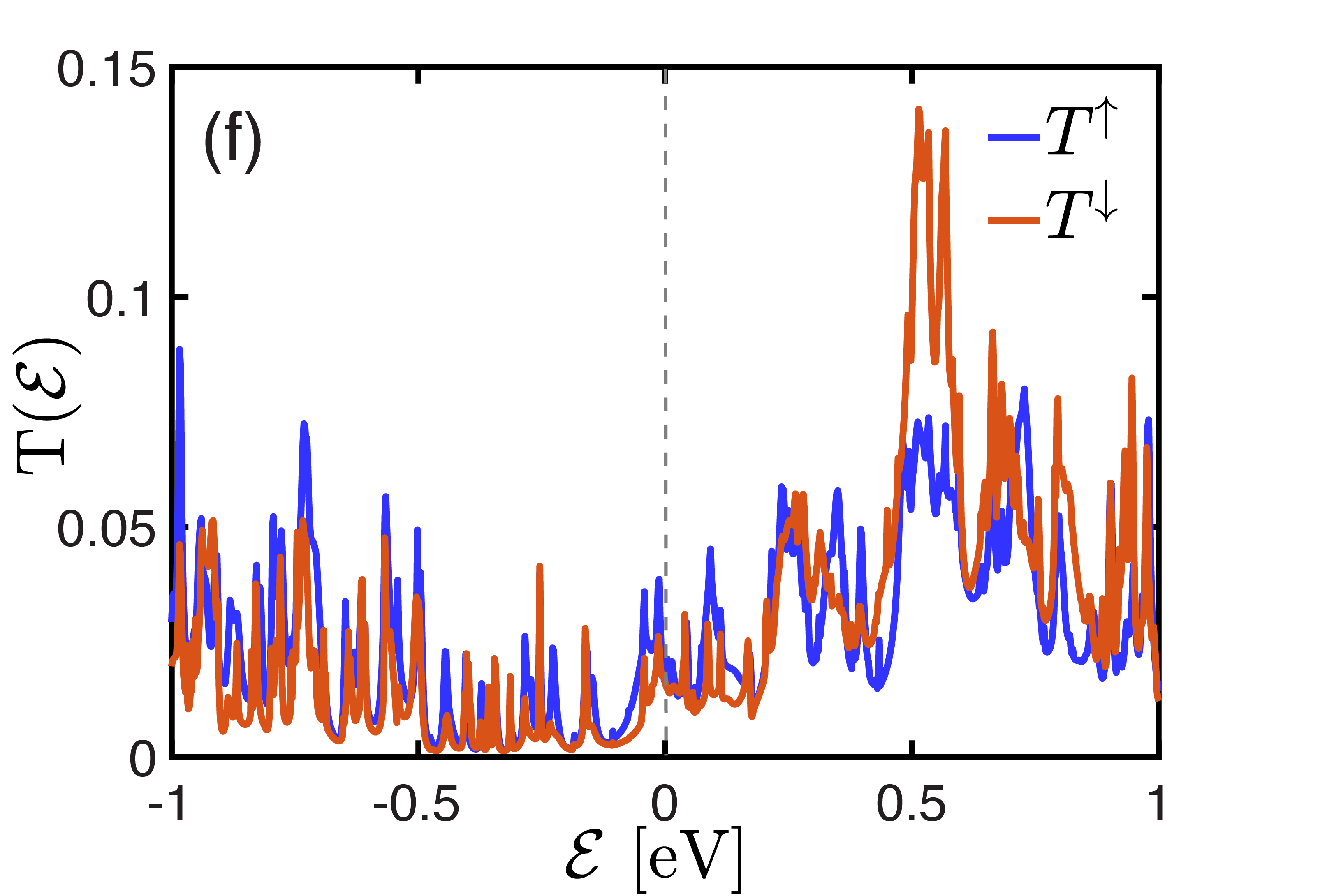}\includegraphics[width=0.23\linewidth]{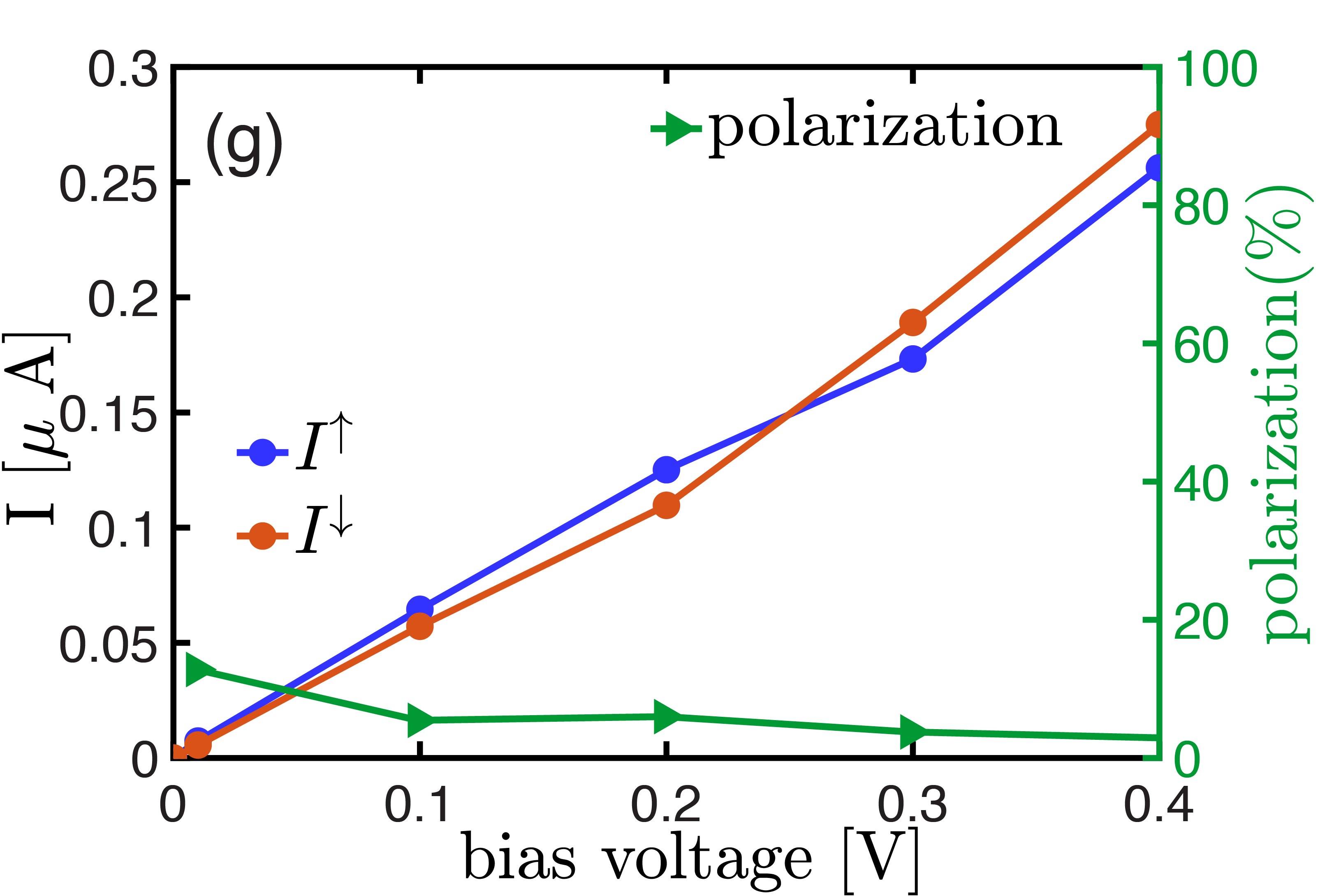}\\
\includegraphics[width=0.23\linewidth]{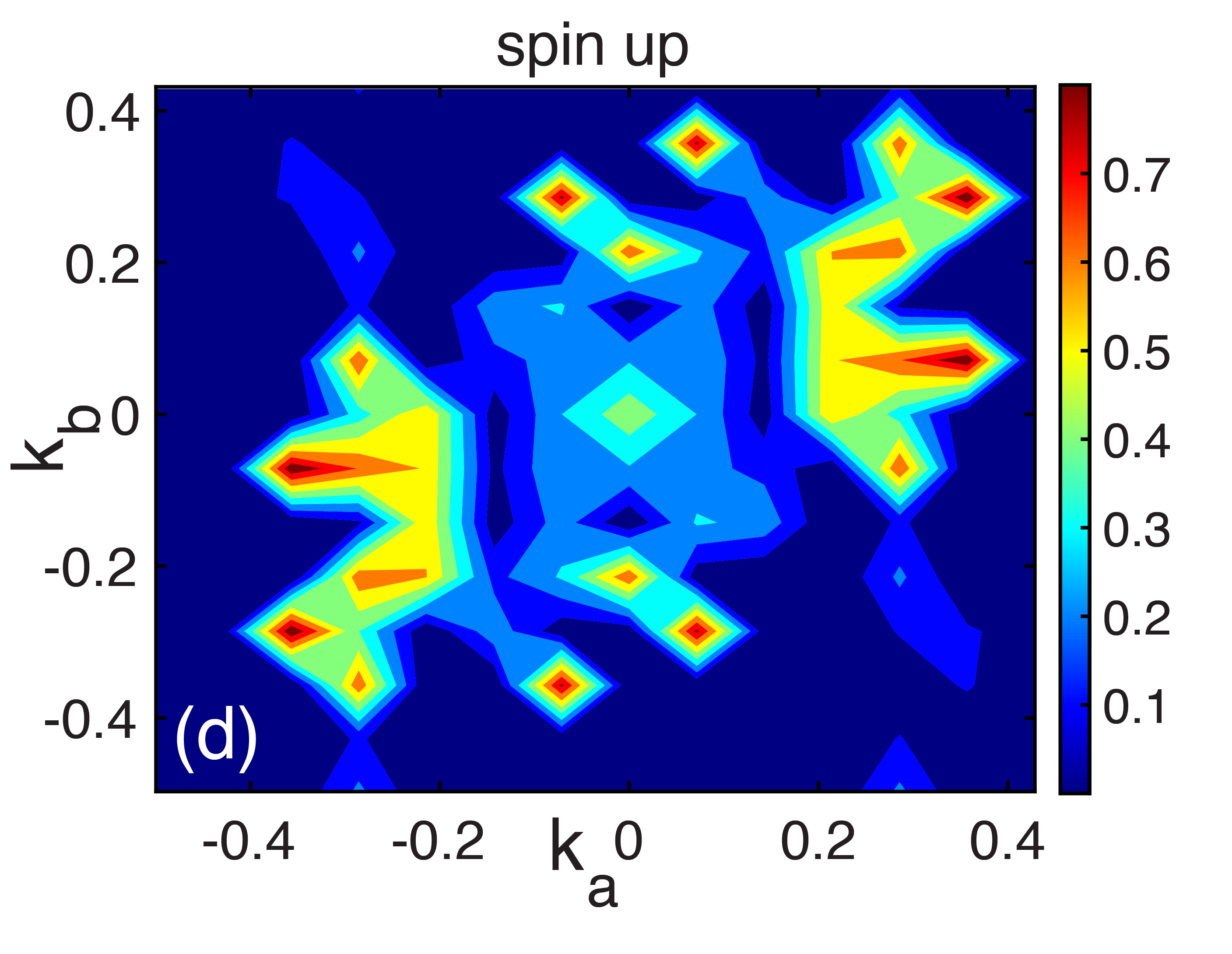}\includegraphics[width=0.23\linewidth]{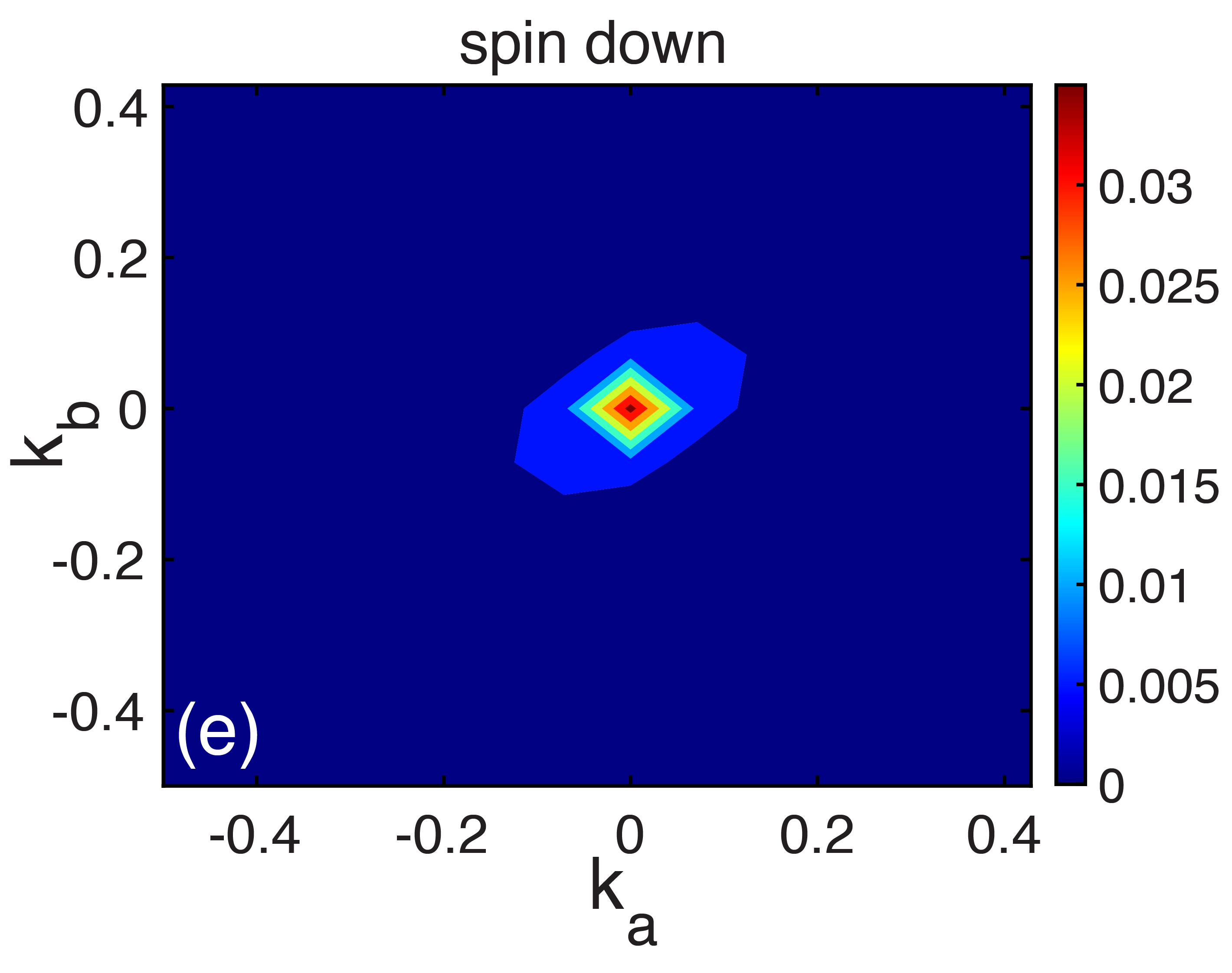}
\hspace{0.7cm}\includegraphics[width=0.23\linewidth]{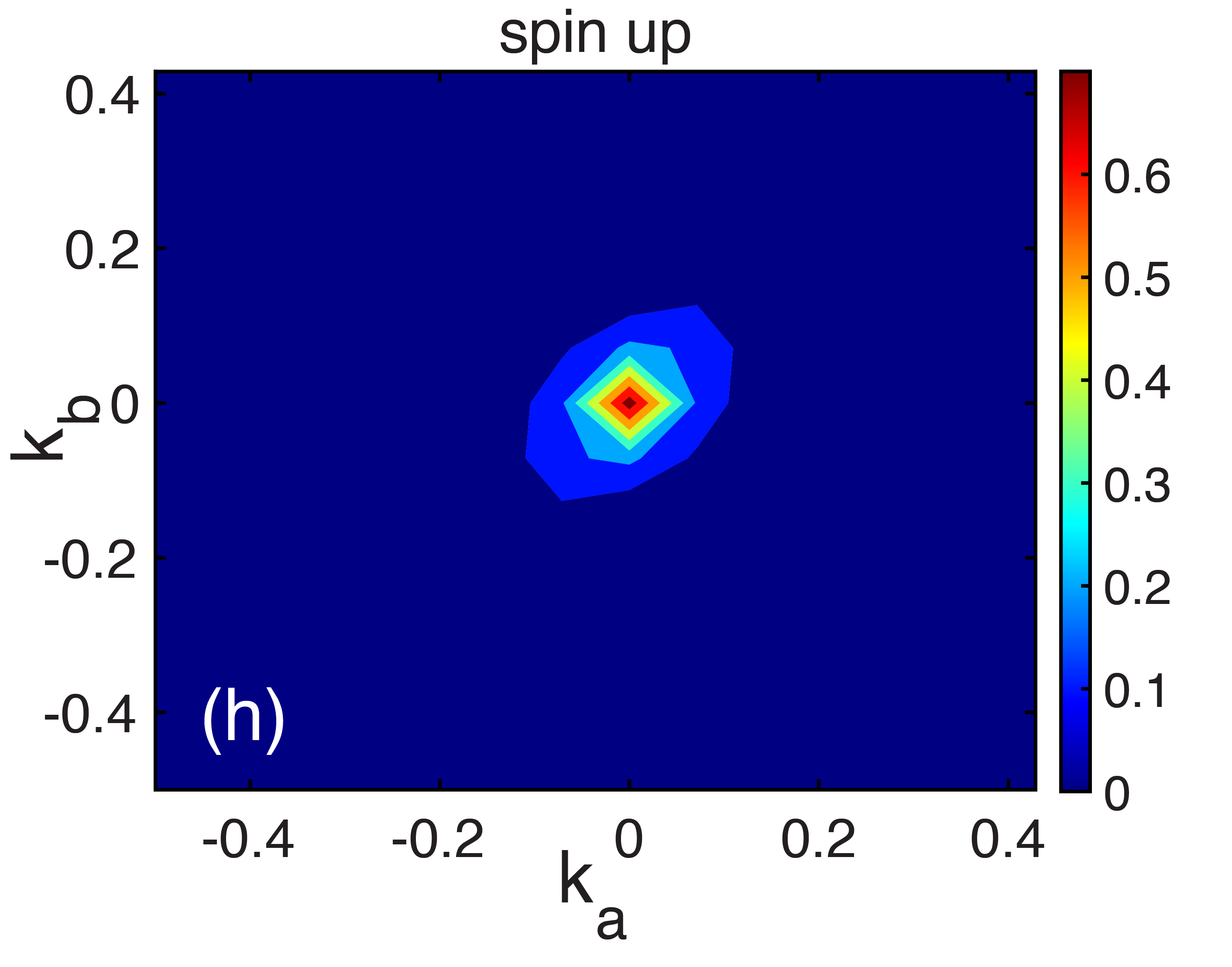}\includegraphics[width=0.23\linewidth]{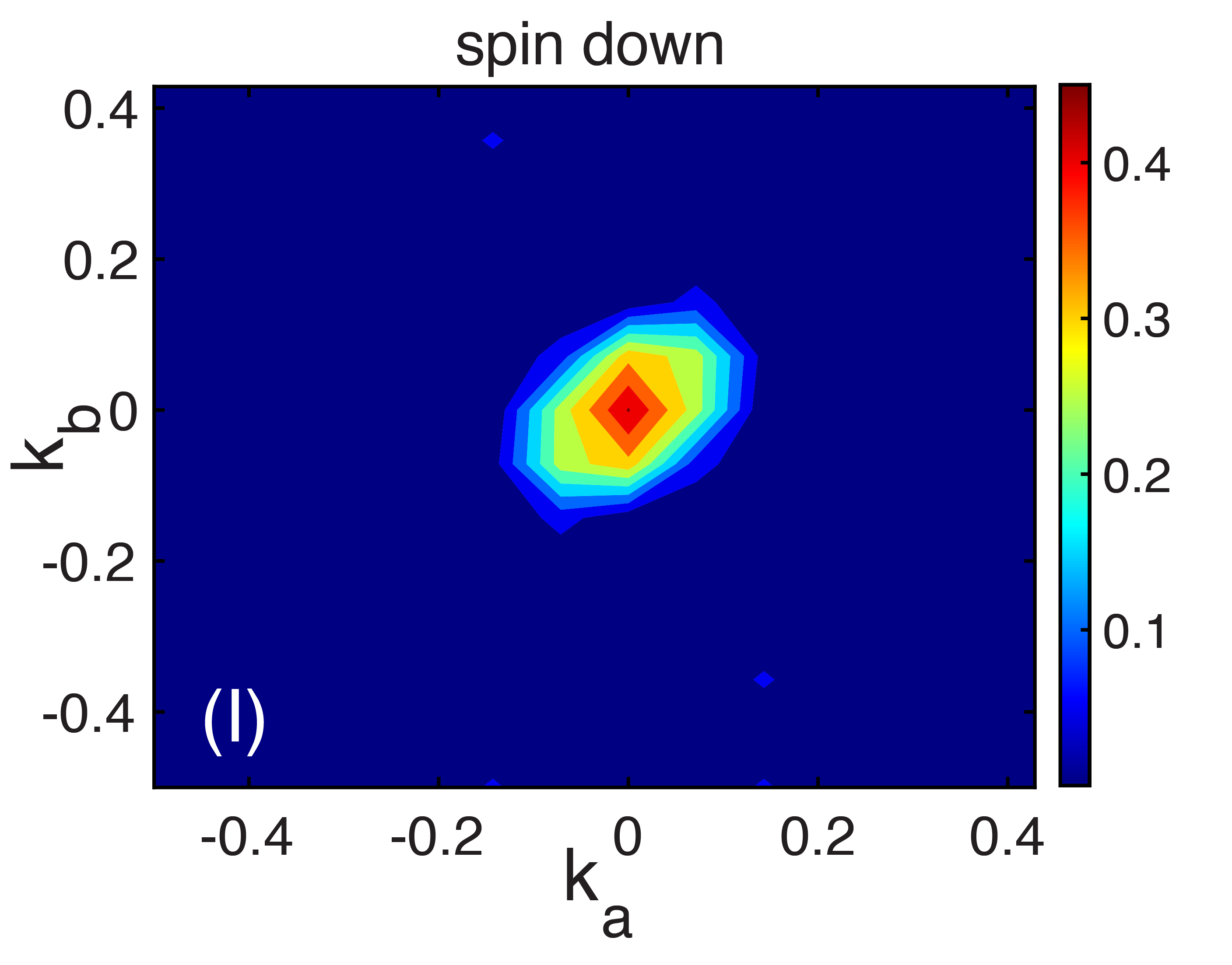}
\caption{\raggedright (a) Schematic view of the PtTe$_2$/bilayer F4GT/PtTe$_2$ heterostructure. Zero-bias spin-dependent transmission coefficient of the system with  (b)FM and (f)AFM configuration. 
I-V characteristics and spin polarization of the heterostructure for FM and AFM device configuration are shown in panels (c) and (g), respectively. 
Zero bias $k_\parallel$-resolved transmission probability at the Fermi energy for (d,h) spin-up and (e, I) spin-down for the system with (d, e) FM and (h, I) AFM configuration.}
    \label{TE_bi}
\end{figure*}
Figure~\ref{f_TEmono} presents the transmission spectrum of a monolayer F4GT sandwiched between two PtTe$_2$ electrodes (panel a) under zero bias voltage. The results show ballistic transport near the Fermi level with distinct spin polarization (Fig.~\ref{f_TEmono}b). In the vicinity of $\mathcal{E}_F$, the transmission coefficient for the spin-up channel is higher, while the spin-down channel exhibits a broad transmission peak reaching its maximum value of 0.75 at $\mathcal{E}_F$+0.6 eV. In contrast, the transmission coefficient for the spin-up channel remains comparatively lower at this energy level. 
This behavior is also supported by the DOS plot in Fig.~\ref{fDOS}, where a peak for the spin-down channel is observed at the same energy level ($\approx$ $\mathcal{E}_F$+0.6 eV). This peak results in a higher contribution of the spin-down channel to the transport properties of the system. 
The difference in the transmission behavior between the two spin channels strongly indicates the presence of spin-polarized transport properties in the F4GT-based device configuration. This can be seen in Fig.~\ref{f_TEmono}c, which exhibits a non-zero spin-polarized current for both spin channels. The I-V curve shows that the spin-up channel has a higher value of current, indicating a preferential flow of electrons with specific spin orientations. Similar spin filtering behavior in F3GT has been reported in conjunction with Cu electrodes~\cite{lin}.

\begin{figure}[b]
\centering
\includegraphics[width=0.89\linewidth]{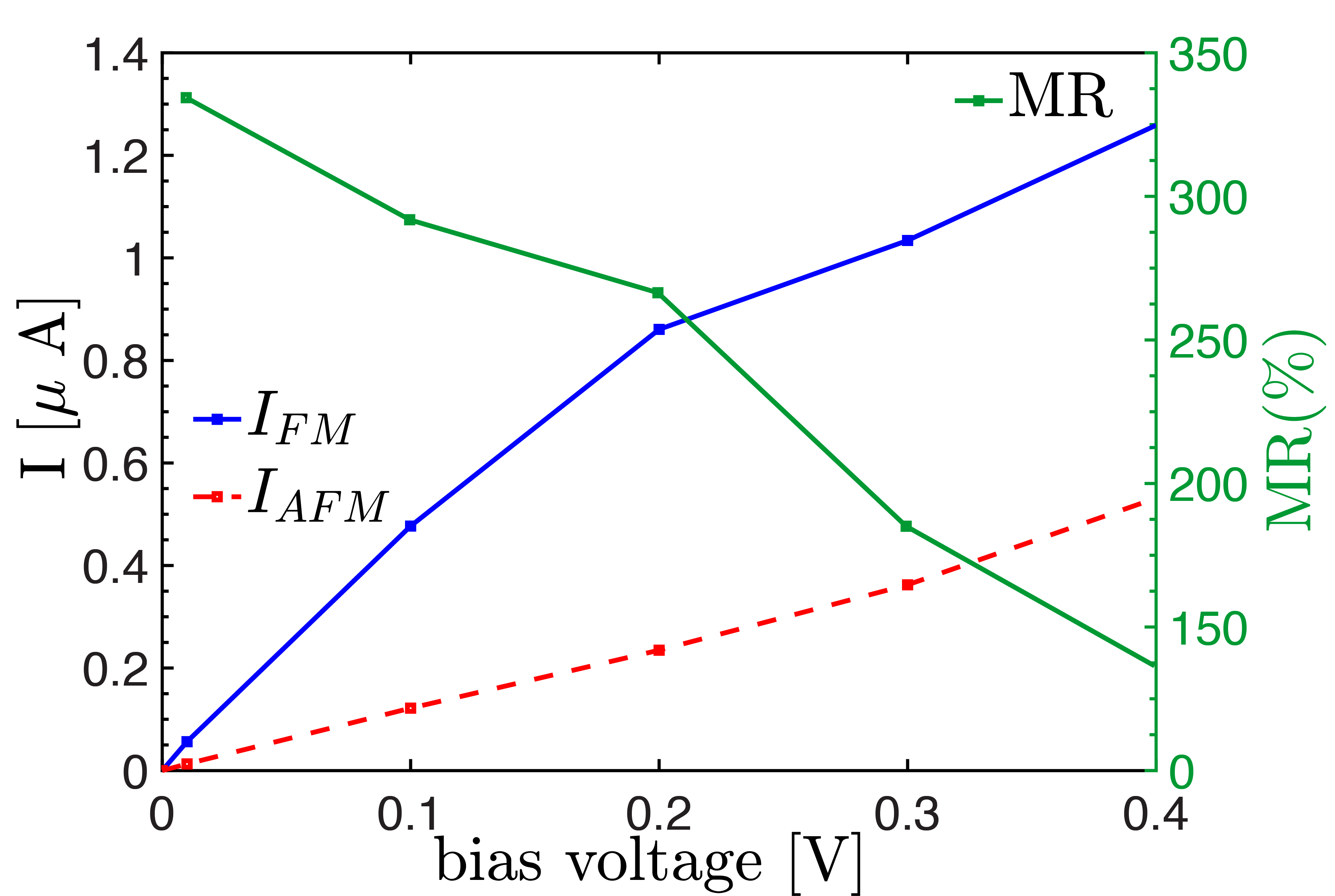}
\caption{\raggedright Total current and variation of the MR for PtTe$_2$/(BL)F4GT/PtTe$_2$ system with FM and AFM configurations as a function of bias voltage.}
    \label{MR}
\end{figure}

Additionally, we have conducted calculations to determine the spin polarization of the current, which is defined as $P=|I_\uparrow-I_\downarrow| /(I_\uparrow+I_\downarrow )$.
Since we require a finite bias voltage for polarization calculations, we have included polarization plots for all cases starting from 0.01 V throughout the paper.
In Fig.~\ref{f_TEmono}c, the current spin polarization exhibits an impressive value of 80$\%$, which remains consistently high in the voltage range of 0.1–0.4 V, comparable to the polarization observed at very low bias voltages. However, as the bias voltage is further increased to 0.5 V, the polarization gradually decreases and reaches 64$\%$. This indicates that single-layer F4GT serves as an effective material for achieving substantial spin polarization, particularly at lower bias voltages. The significant transport polarization observed in F4GT agrees well with experimental results. A very recent study employing spin-resolved Andreev reflection spectroscopy on F4GT revealed an exceptionally high transport spin polarization, surpassing 50$\%$~\cite{PhysRevB.107.224422}.

The $k_\parallel$-resolved transmission probability offers a comprehensive understanding of how electrons with different spin orientations propagate through the material at the Fermi level. 
Thus, in Fig.~\ref{f_TEmono}, we performed calculations to determine the momentum-dependent transmission for Fig.~\ref{f_TEmono}(d) spin-up and Fig.~\ref{f_TEmono}(e) spin-down electrons at the Fermi energy under zero bias voltage.   
Here, the $k_\parallel$-resolved transmission probability utilizes the vectors $k_a$ and $k_b$, as outlined in Fig.~\ref{f1}d. This particular choice of vectors leads to a square-shaped representation for the $k_\parallel$-resolved transmission probability plots, despite the hexagonal unit cell of F4GT. 
As expected, the transmission probability for spin-up electrons surpasses that of spin-down electrons, indicating a significant degree of spin polarization in the material. 
Furthermore, we found that the transmission in both the spin-up and spin-down channels is independent of the in-plane wave vector ($k_\parallel$) direction indicating that the transmission does not vary with the rotation of the $k_\parallel$. This is consistent with the structural symmetry of the system.

Quantum transmission eigenstates, which characterize electron propagation within a device, are depicted via isosurface plots in Fig.~\ref{eigenTE}. The figure shows a monolayer F4GT placed between two PtTe$_2$ electrodes for both (a) spin-up and (b) spin-down channels at the Fermi energy level under zero bias. A quantum transmission eigenstate can be thought of as a combination of two distinct electron states. One of these states represents electrons moving from the left electrode to the right electrode, while the other describes electrons going from the right electrode to the left electrode. Their relative phase depends on their proximity to the respective electrodes.
This phase difference results in an interference-like pattern in the isosurface plot, particularly in the PtTe$_2$ layers on the left and right sides of F4GT, far from the scattering region. A $\pi$ phase shift is observed for the transmission eigenstates localized on the left electrode in the spin-down channel (panel b) compared to the spin-up channel (panel a). Nevertheless, for the PtTe$_2$ layers on the right side of F4GT, the phase of eigenstates is the same for both channels. In Fig.~\ref{eigenTE}(a), the transmission eigenstate in the scattering region exhibits a pattern characterized by $d_z^2$ orbitals on Fe atoms for the spin-up channel. However, for the spin-down channel in Fig.~\ref{eigenTE}(b) this pattern displays a greater degree of d-p hybridization states. This hybridization arises from the increased contribution of p orbitals from Te atoms, interacting with the Fe $d_z^2$ orbitals.
Interestingly, we observe that in the right electrode region, the transmission eigenstates pertaining to the spin-up channel (panel a) exhibit greater dominance when compared to those of the spin-down channel (panel b). 
A comprehensive examination of the fundamental mechanisms contributing to this spin-filtering phenomenon reveals that the dominant transmission eigenstates in the spin-up channel establish a robust transmission channel in the heterojunction. This enables the efficient movement of electrons from the left electrode to the right electrode. Conversely, the spin-down channel experiences a relative scarcity of transmission eigenstates (Fig.~\ref{eigenTE}b), resulting in a restricted transmission of electrons to the right side.
The observed spin-filtering effect originates from the channel-selective transmission behavior, where the spin-up channel displays a more pronounced transmission, allowing a larger number of electrons to traverse from the left to the right side.
The orbital-projected local density of states is presented in Fig.~\ref{ldos} for distinct orbitals, focusing on (a) the left electrode's PtTe$_2$ layer interfacing with F4GT, (b) F4GT itself, and (c) the right electrode's PtTe$_2$ layer interfacing with F4GT. The figure highlights a substantial contribution from both p and d orbitals, resulting in discernible d-p hybridization states near the Fermi level. Notably, in panel b, it is evident that the contribution of the d orbital in the spin-down channel is more than the spin-up channel, indicating a higher d-p hybridization states in the spin-down channel , aligning with the observations in Fig.\ref{eigenTE}d.

\begin{figure*}[t]
    \centering
    \includegraphics[width=0.4\linewidth, trim={0cm -3cm 0cm 0.4cm}, clip]{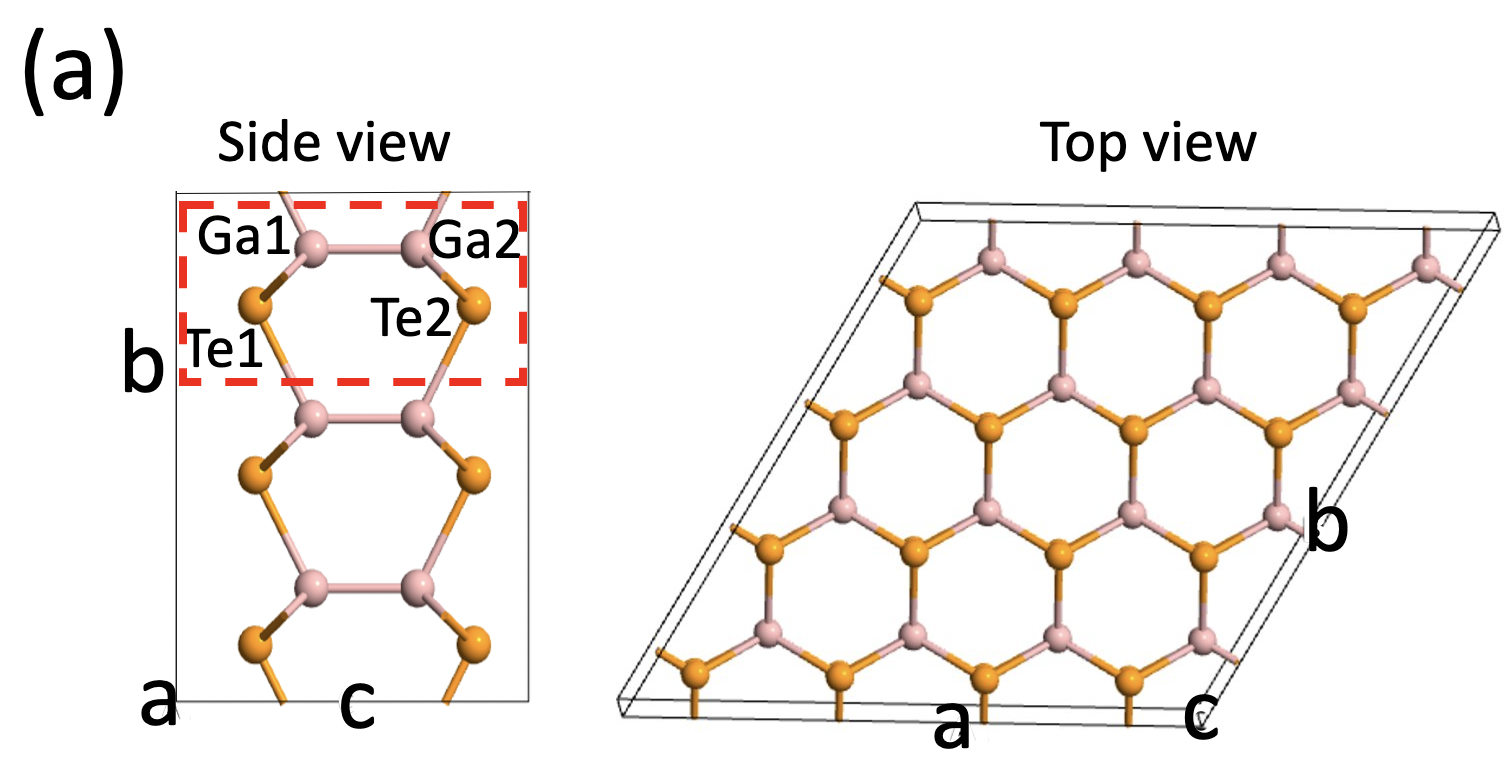}\hspace{0.6cm}\includegraphics[width=0.35\linewidth]{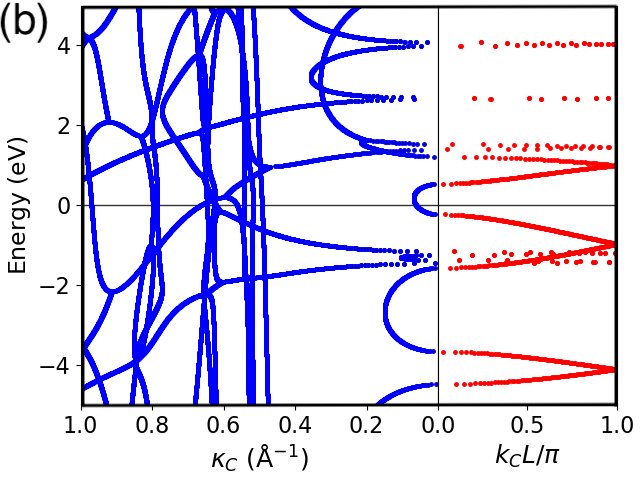}\\
    \includegraphics[width=0.33\linewidth]{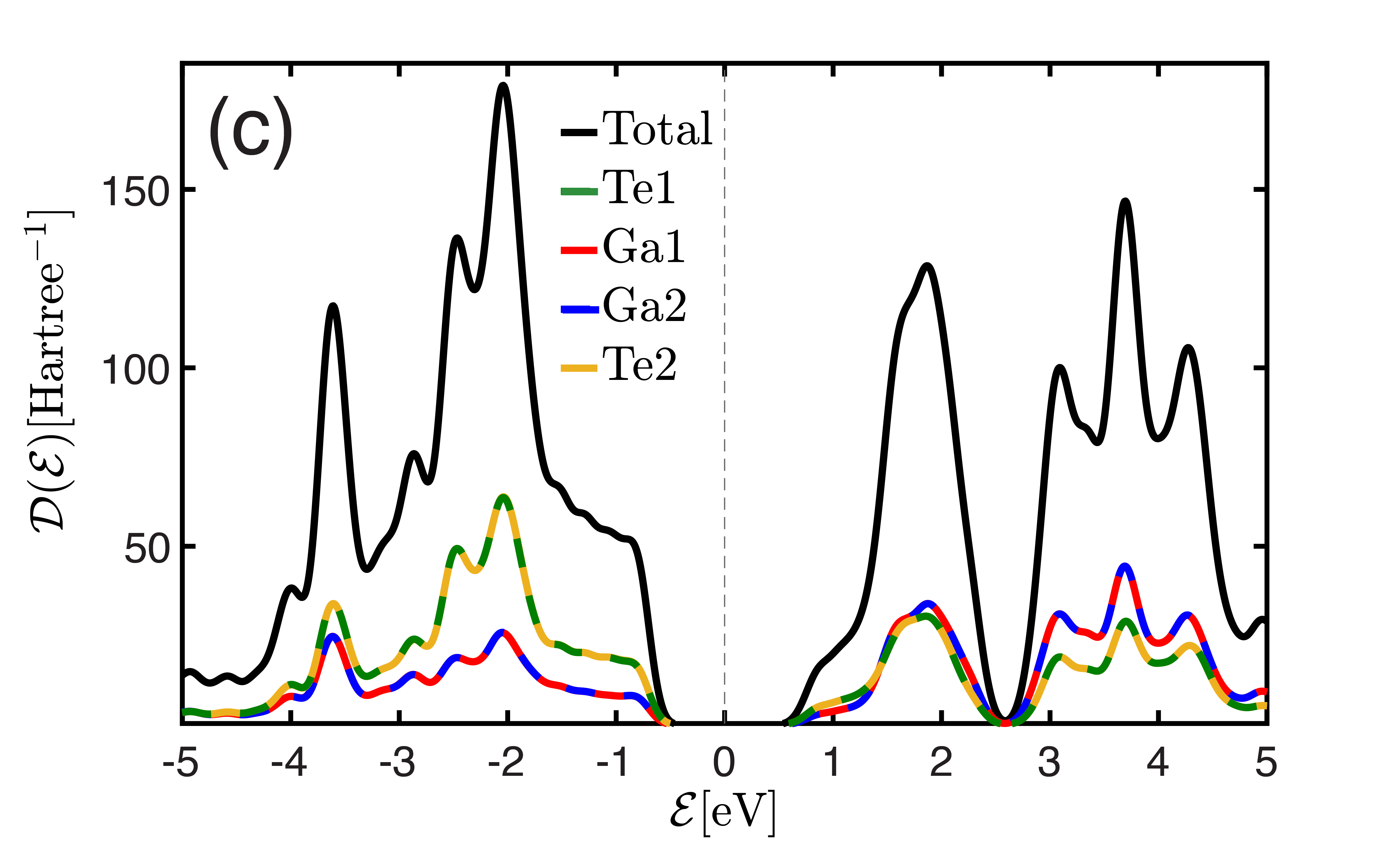}\includegraphics[width=0.33\linewidth]{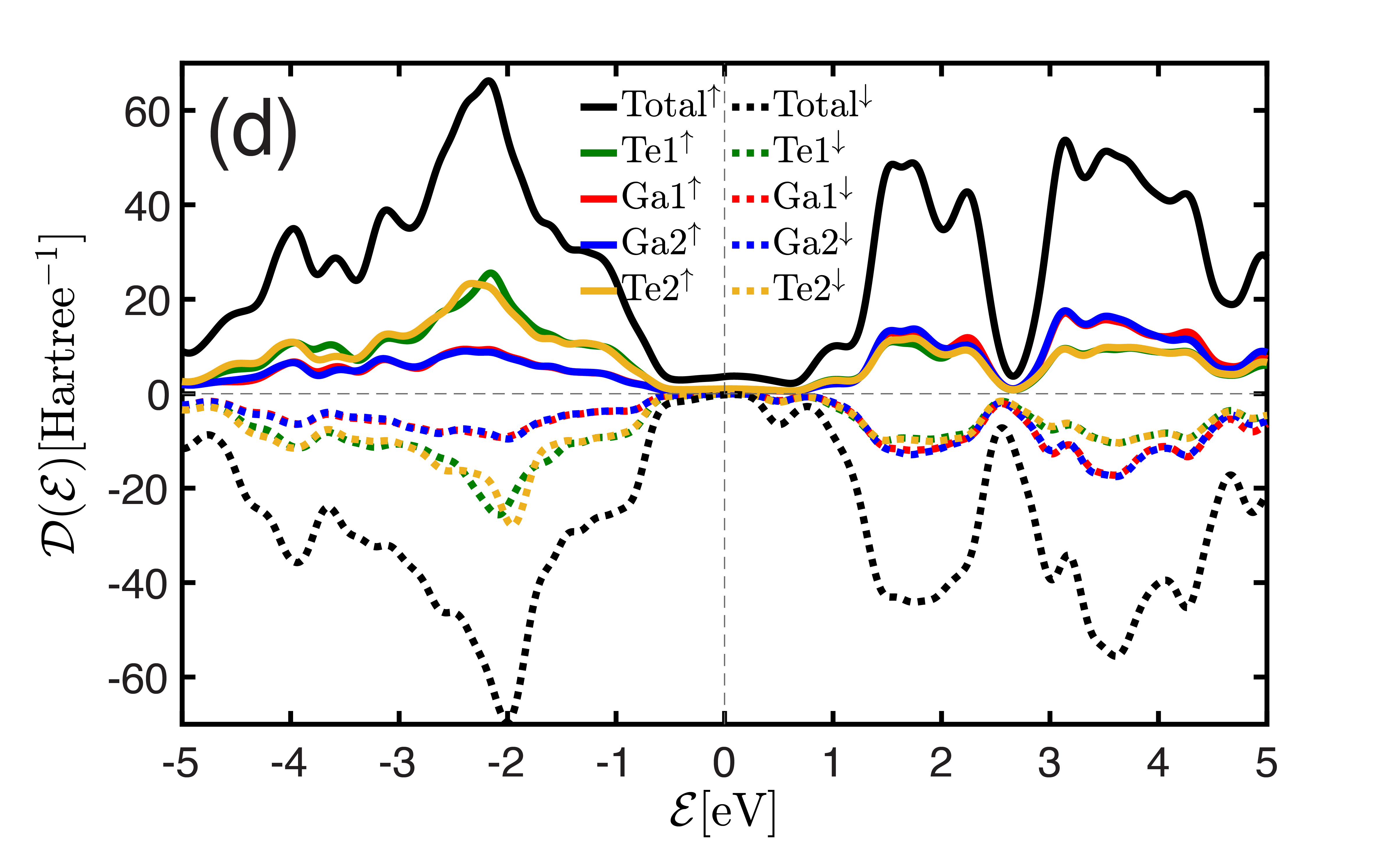}\includegraphics[width=0.33\linewidth]{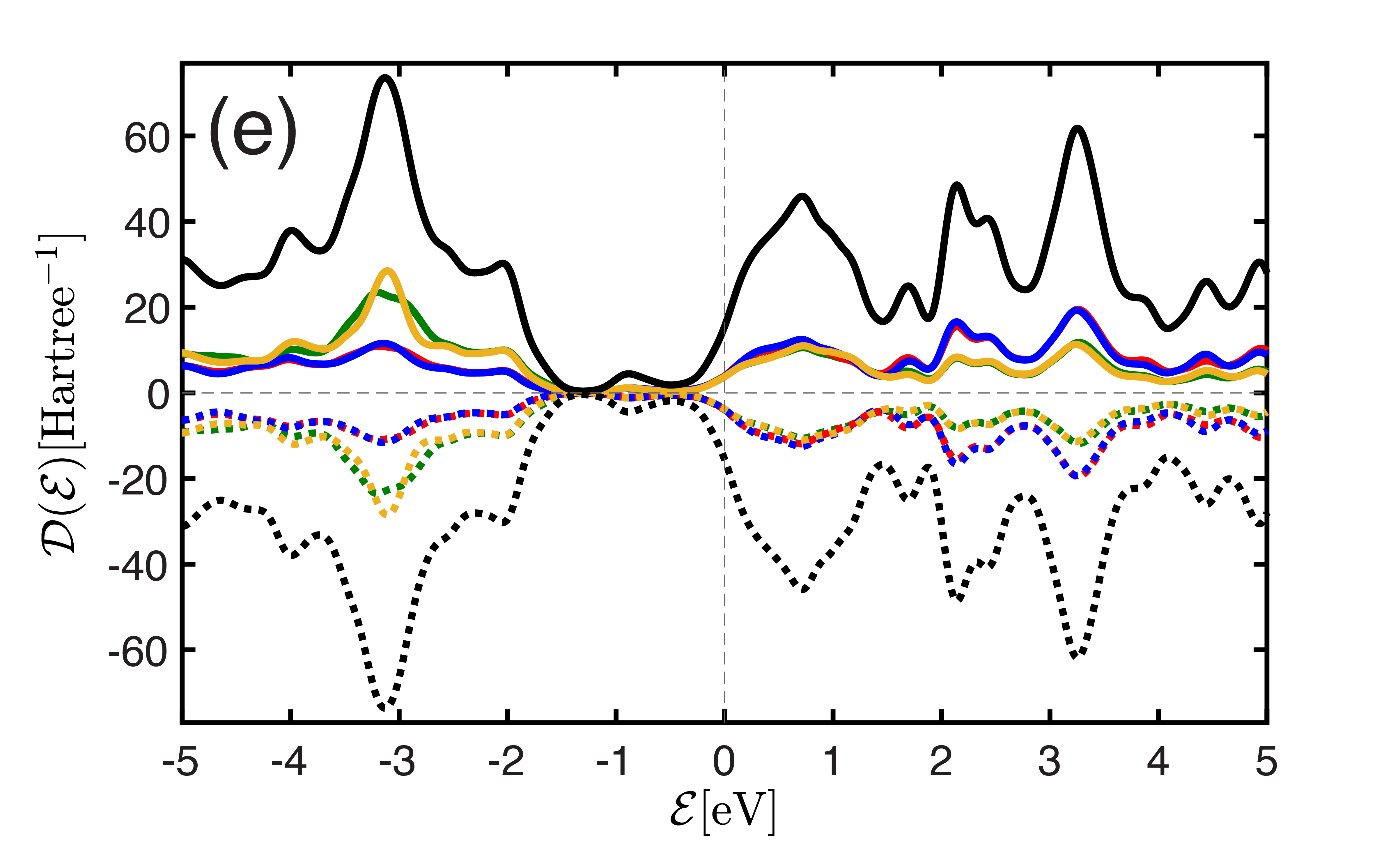}
    \caption{(a) Side and top views of the atomic structure of monolayer GaTe. Dashed rectangular shows its unit cell. 
    (b) Calculated complex band structures of bulk GaTe, $L$ is the layer separation perpendicular to the cleave plane. Both the real bands (right panel) and imaginary bands (left panel) are plotted. Projected density of states for (c) freestanding monolayer GaTe, and monolayer GaTe in the $PtTe_2/F4GT/GaTe/F4GT/PtTe_2$ heterostructure for (d) parallel and (e) antiparallel configurations.}
    \label{dos-GaTe}
\end{figure*}

Furthermore, we have calculated the spin-polarized electronic structure and transport properties for the bilayer Fe4GT placed between two PtTe$_2$ electrodes. The AB stacking for two F4GT layers is selected due to its higher stability in terms of energy compared to the AA stacking configuration.
The two-probe model for the Non-Equilibrium Green's Function (NEGF) calculations of this bilayer system is illustrated in Fig.~\ref{TE_bi}a. Figure \ref{TE_bi}b displays the spin-resolved energy-dependent transmission function of the system at zero voltage, with the energy measured relative to the Fermi level and the bilayer Fe4GT in a ferromagnetic (FM) configuration. Similar to the monolayer system, the zero-bias ballistic transport near the Fermi level exhibits spin polarization. However, in the bilayer system, there is a broad transmission peak between approximately [$\mathcal{E}_F$-0.1, $\mathcal{E}_F$+0.1] eV, with its maximum at $\mathcal{E}_F$ for the spin-up channel. In contrast, the transmission coefficient in the spin-down channel is suppressed. This behavior results in a spin filtering effect, allowing only one spin type to flow through the constriction at this particular energy range.
The I-V calculation (Fig.~\ref{TE_bi}c) reveals that the difference in spin-up and spin-down currents is more pronounced in the bilayer system compared to the single-layer F4GT, leading to higher polarization of the current. 
This value of spin polarization surpasses the reported spin polarization values for a device comprising a bilayer of F3GT placed between Cu electrodes. The highest spin polarization value for the Cu/F3GT/Cu heterostructure with a ferromagnetic configuration is documented as 85$\%$ at a very low bias voltage~\cite{lin}. The significantly enhanced spin polarization in our FM system indicates its potential for efficient spin transport in spintronic devices.

To gain further insight into the transport properties of the systems, we analyzed the distribution of transmission coefficients at the Fermi level under zero bias voltage, as depicted in Fig.~\ref{TE_bi}(d, e). In comparison to the spin-down channel shown in panel (e), the $T(\mathcal{E})$ for the spin-up channel in panel (d) exhibits higher values and is distributed across most of the first Brillouin zone region. However, for the spin-down channel, the transmission contours are primarily concentrated around the $\Gamma$ point. This indicates that the spin-up electrons have a significantly higher probability of transmission, while the transmission of the spin-down electrons is largely suppressed.

We also explored the transport properties of the bilayer Fe4GT system with an antiferromagnetic (AFM) configuration sandwiched between two PtTe$_2$ electrodes. 
Our calculations reveal that the total energy of the FM system is remarkably close to that of the AFM configuration, differing by only 0.016 eV. The small difference between the total energy indicates that these two configurations are nearly degenerate in energy. In other words, both FM and AFM states are stable and energetically favorable.
As illustrated in panel (f) of the figure, the zero-bias transmission spectrum reveals lower transmission compared to the FM case and both the spin-up and spin-down channels exhibit similar behavior near the Fermi level. The I-V characteristics in panel (g) indicate that less current passes through the device with the AFM configuration compared to the FM one.
Notably, both spin-up and spin-down channels show the same I-V curve, implying that there is no significant spin-filtering effect for this device. Additionally, the spin polarization curve confirms very low spin filtering in the system, ranging from 14$\%$ to 3$\%$ for the given bias voltage range of 0.01 V to 0.4 V.  

Additionally, the $k_\parallel$-resolved conduction map in panels (h) and (i) for spin-up and spin-down channels in the PtTe$_2$/(BL)F4GT/PtTe$_2$ system with AFM configuration displays isotropic transmission probability, primarily concentrated around the $\Gamma$ point. 

The total current, a summation of both spin-up and spin-down currents, is depicted in Fig.~\ref{MR} for the PtTe$_2$/(BL)F4GT/PtTe$_2$ system, which includes both FM and AFM configurations, as a function of bias voltage. As expected, the AFM state reveals a significant decrease in the transmission of both spin-up and spin-down electrons, resulting in a reduced total current compared to the FM configuration.
The Magnetoresistance (MR) ratio is computed using the formula MR $= (1/I_{AFM} - 1/I_{FM}) / (1/I_{FM})$. The MR plot is illustrated in Fig.~\ref{MR} where it is observed at low bias levels, the system demonstrates remarkable MR of 334$\%$, signifying efficient magnetoresistive behavior. However, as the bias voltage increases, the MR gradually diminishes.

\begin{figure*}[t]
\centering
\includegraphics[width=.75\linewidth]{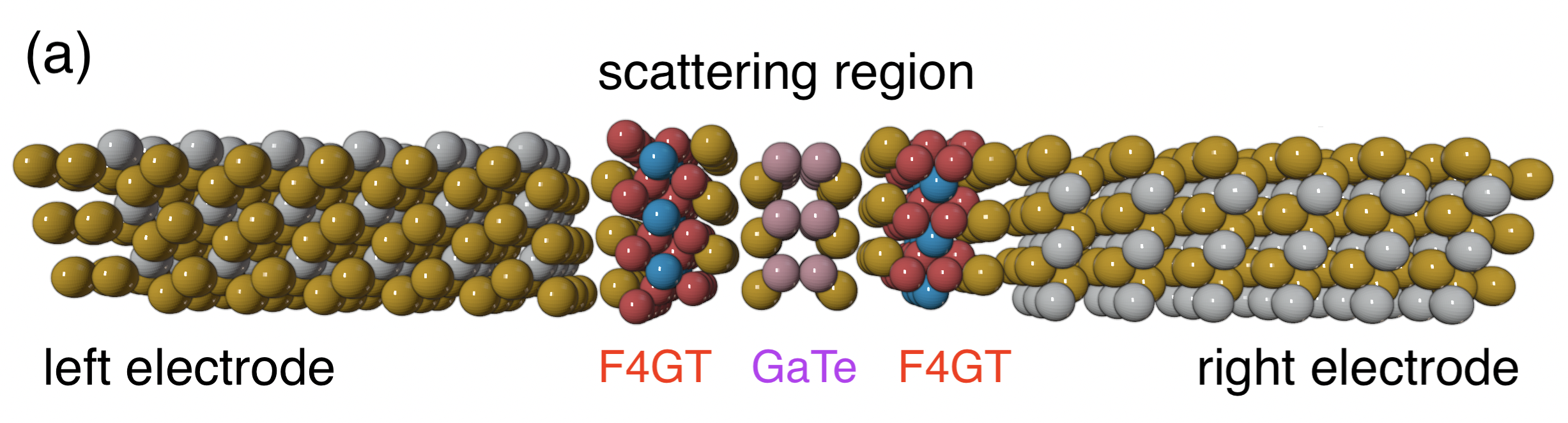}\\
{\hspace{1cm}\large\raggedleft{\textbf{Parallel }}\hspace{7cm}\raggedleft{\textbf{Antiparallel }}} \vspace{0.1cm}\\
\includegraphics[width=0.33\linewidth, trim={0cm -0.3cm 0cm 1cm}, clip]{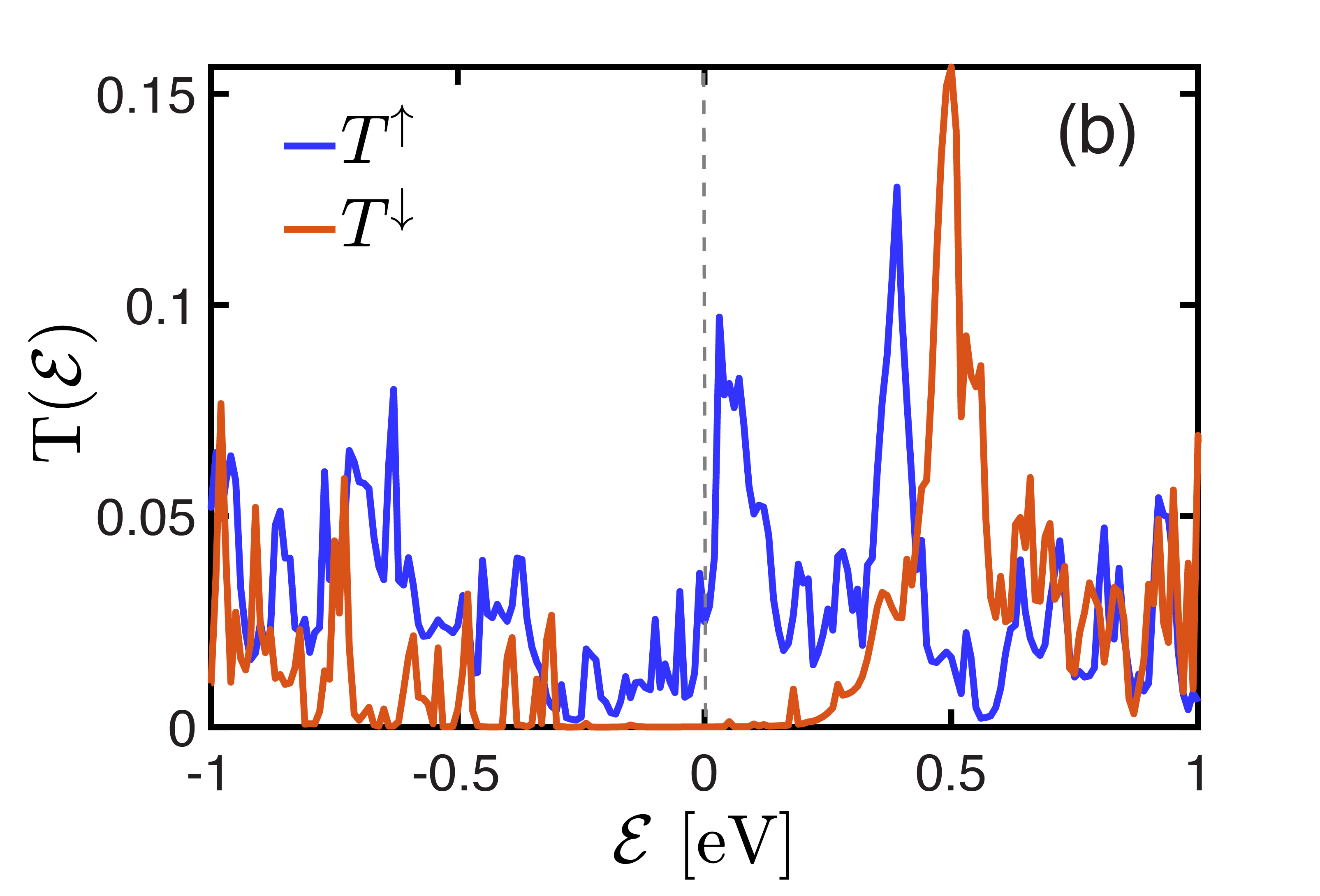}
\hspace{3cm}\includegraphics[width=0.33\linewidth, trim={0cm -0.3cm 0cm 0.78cm}, clip]{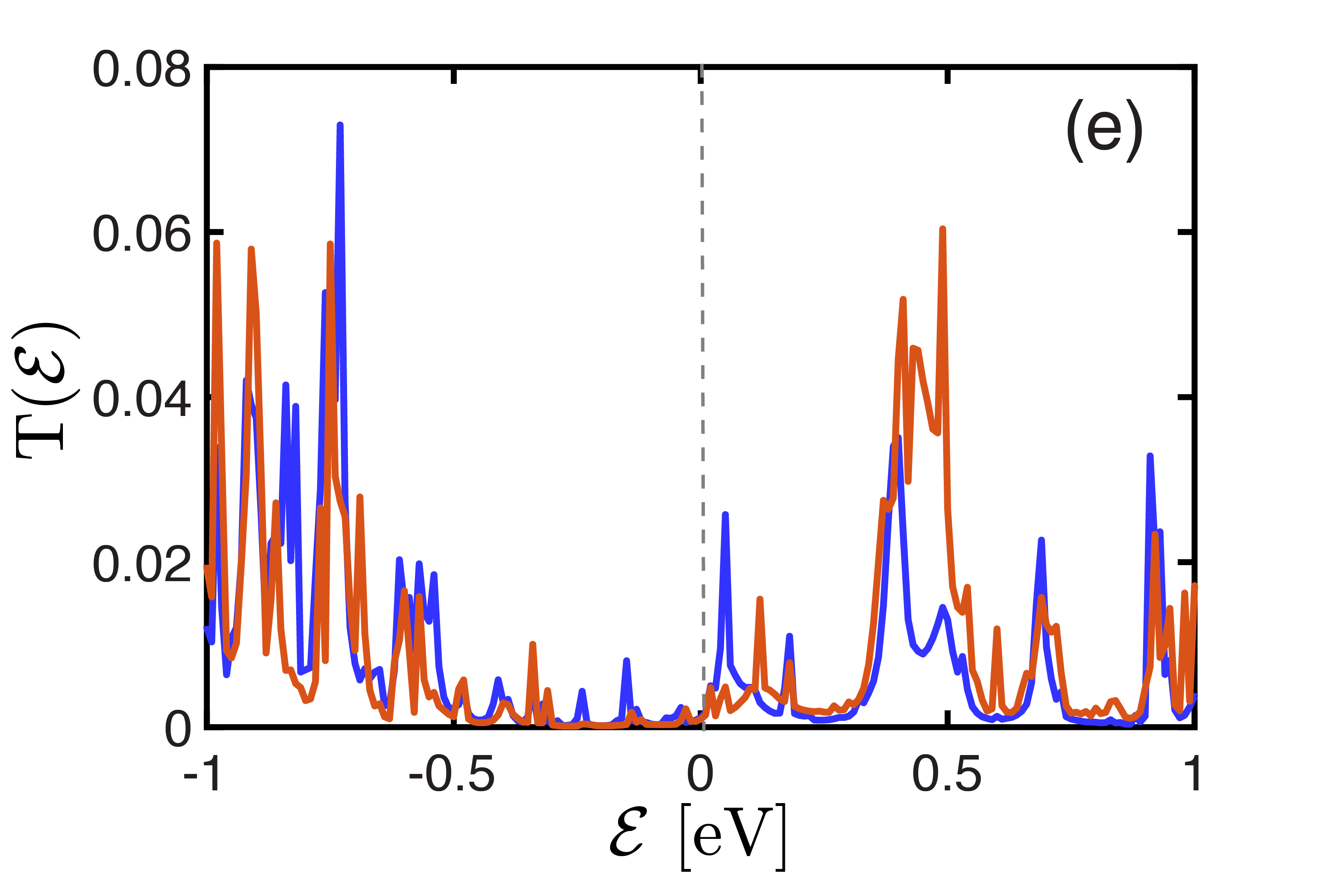}\\
\includegraphics[width=0.24\linewidth]{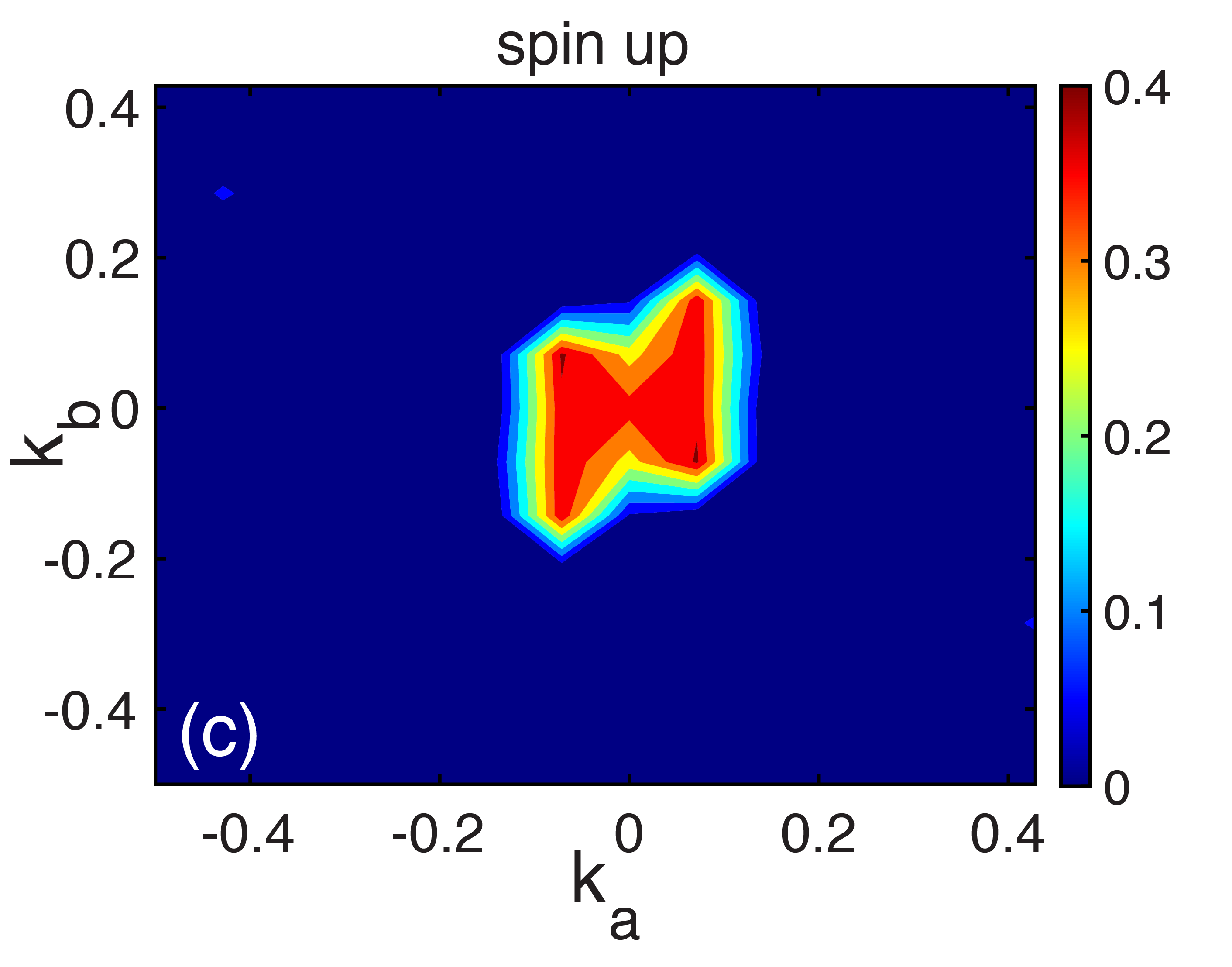}\includegraphics[width=0.24\linewidth]{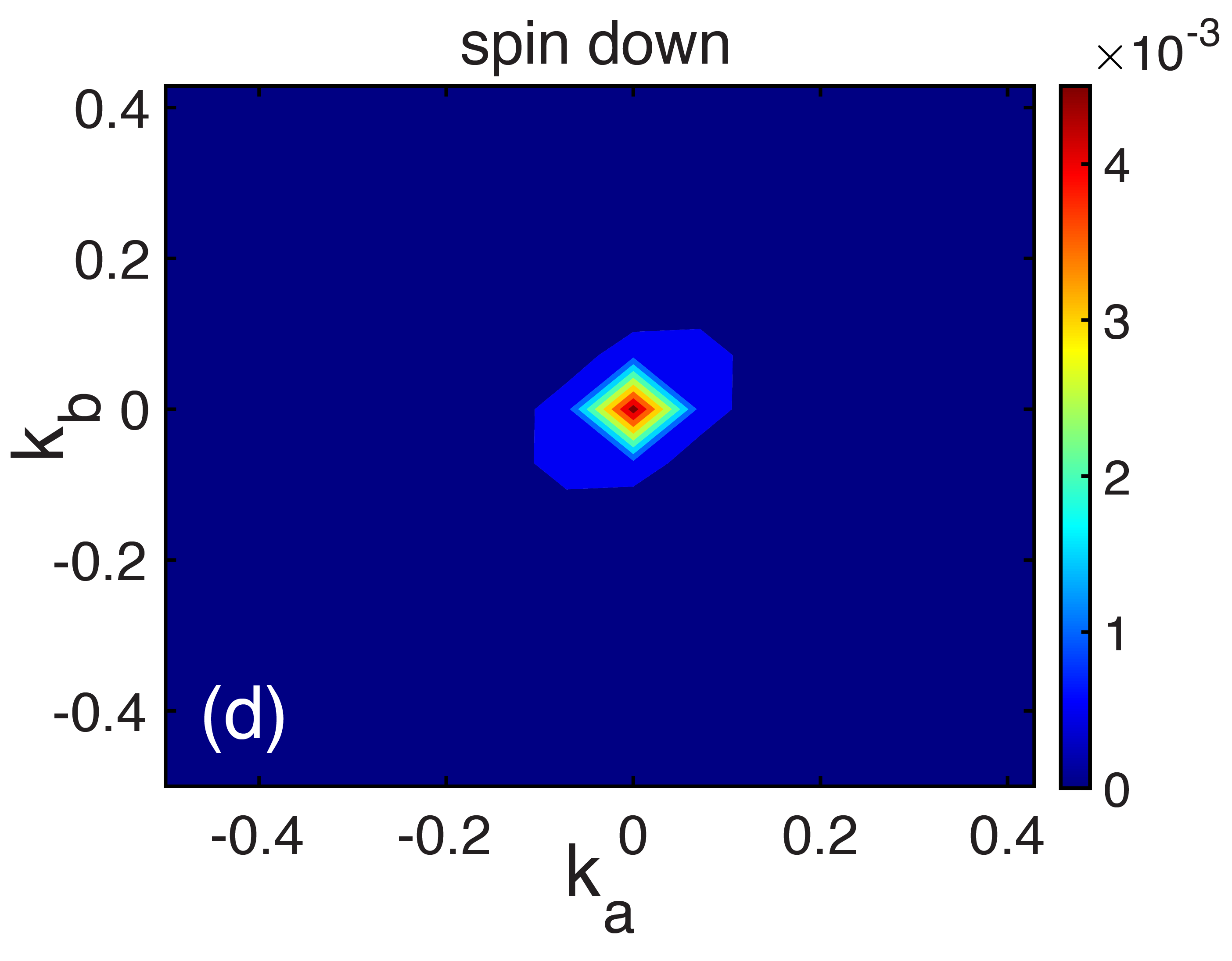}
\hspace{0.5cm}\includegraphics[width=0.24\linewidth]{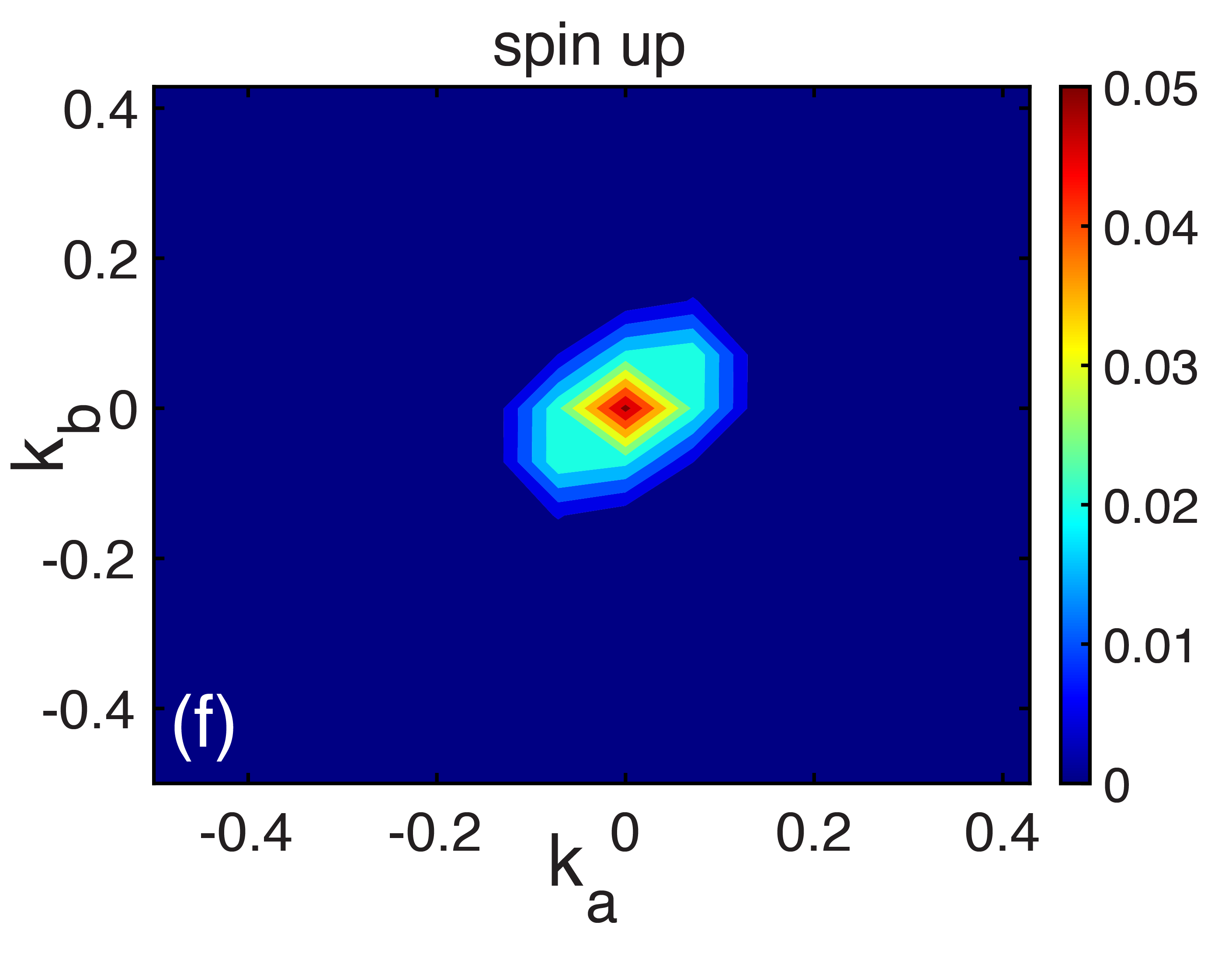}\includegraphics[width=0.24\linewidth]{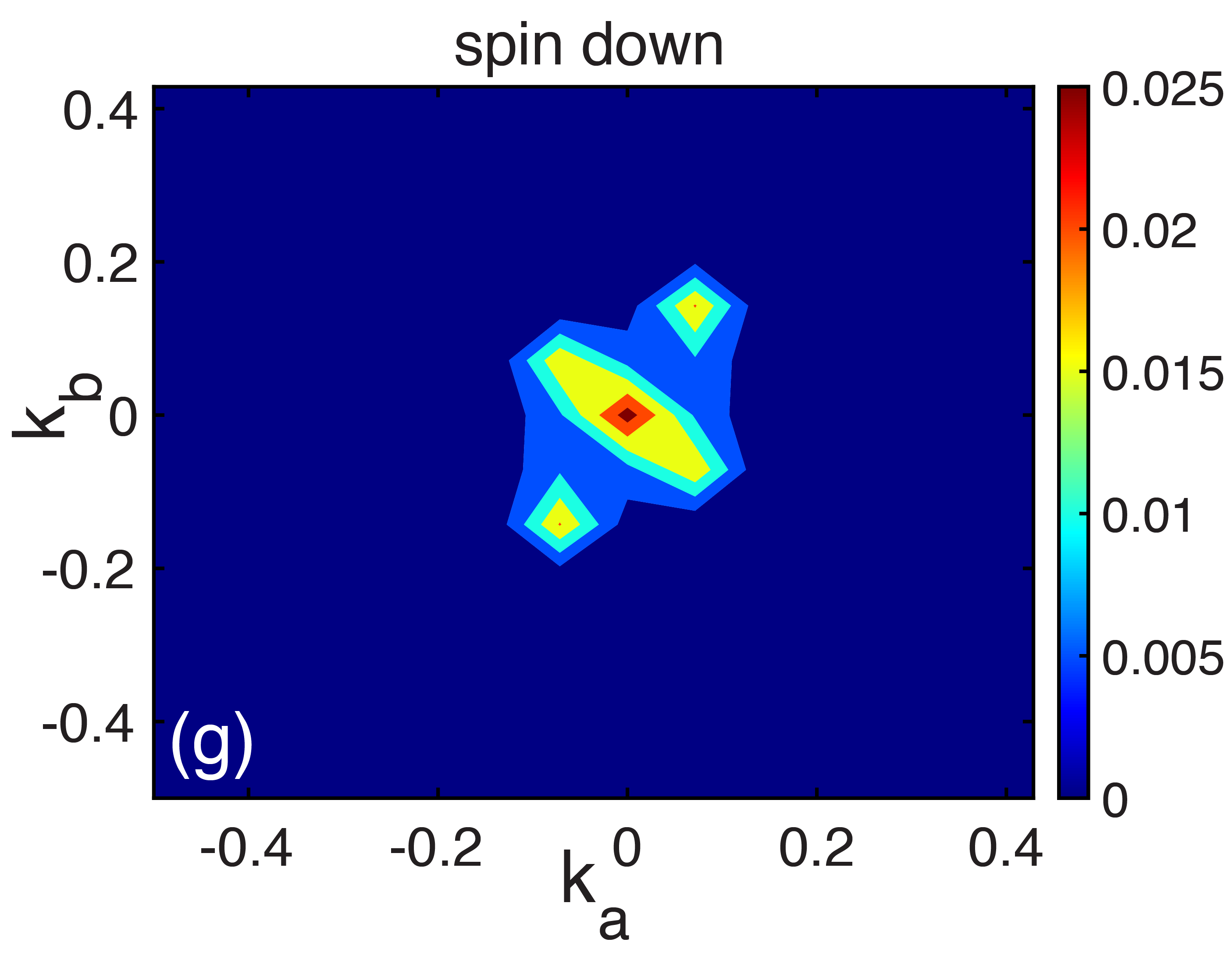}
\caption{\raggedright (a)Two-probe model used for NEGF calculations, depicting a single-layer Fe$_4$GeTe$_2$/GaTe/Fe$_4$GeTe$_2$ magnetic tunnel junction sandwiched between two PtTe$_2$ electrodes. Transmission characteristics of the system in the (b) parallel and (e) antiparallel state under zero bias voltage. $k_\parallel$-resolved transmission probability at the Fermi energy for (c,f) spin-up and (d,g) spin-down for (c,d) parallel and (f,g) antiparallel state at zero bias voltage.}
    \label{full_dev}
\end{figure*}

\subsection{Magnetic tunnel junction}
The study of magnetic tunnel junctions (MTJs) is important in spintronics, as they are key components in spin-based devices~\cite{doi:https://doi.org/10.1002/047134608X.W8231}. The structure of an MTJ consists of ferromagnetic layers separated by a tunneling a barrier between them.
The MTJs can come in different sizes, use low energy and could potentially last without wearing out. These properties make the MTJ highly valuable for various applications like magnetoresistive random access memory~\cite{5388774},  magnetic sensors~\cite{10.1063/1.3701277}, hard disk drive etc~\cite{JIANG2023170546}.

In our investigation, we have introduced GaTe as a barrier between the ferromagnetic electrodes. GaTe is a member of the group-VIII metal chalcogenide family and is a semiconductor with an indirect bandgap. It crystallizes in the $P\bar{6}m2$ space group~\cite{PhysRevB.95.115409, PhysRevB.103.165422}, and its lattice parameter for a single layer is approximately 4.09 \AA, closely matching that of F4GT and PtTe$_2$. This close lattice match minimizes material mismatches in the heterostructure, promoting a better interface quality.
Figure~\ref{dos-GaTe}a presents side and top views of monolayer GaTe's atomic structure, consisting of four atoms in its unit cell: two Ga and two Te atoms. 

Accurately evaluating the exponential decay of wave transmission necessitates a comprehensive analysis of the dispersion spectrum, considering both propagating and evanescent wave modes. In Fig.~\ref{dos-GaTe}b, we present the computed complex band structures of bulk GaTe, showcasing the real bands on the right panel and the imaginary bands on the left. Here, $L$ in the imaginary bands represents the set at 17.52 \AA\, for semiconducting GaTe with AB stacking. The right panel of the plot illustrates the real bands, while the left part exhibits the complex bands plotted against the imaginary part, $\kappa_C$. The calculated band gap of ~0.78 eV aligns with findings from Ref.~\onlinecite{Lai2022}. Evanescent states exhibit a characteristic decay length inversely proportional to the imaginary wave vector $\kappa_C$. Thus, our primary focus is on states where $\kappa_C$ remains small within the gap. Notably, the presence of states with increasing $\kappa_C$ indicates a progressive enhancement of the damping or broadening parameter in the complex band structure, signifying the temporal decay of electronic states.

The total and atomic projected density of states for freestanding monolayer GaTe is displayed in Fig.~\ref{dos-GaTe}c, revealing a semiconductor phase with a band gap of approximately 1.05 eV. Notably, the projected density of states for atoms of the same type within the unit cell exhibits similar behavior.
To investigate changes in the electronic structure of monolayer GaTe when integrated into the $PtTe_2/F4GT/GaTe/F4GT/PtTe_2$ heterostructure, we computed the spin-polarized projected density of states for both parallel and antiparallel configurations. These results are presented in panels (d) and (e) of Fig~\ref{dos-GaTe}, respectively. Our findings indicate that the incorporation of GaTe monolayer in device heterostructure leads to a transformation of its electronic phase from semiconductor to metal, as evidenced by the non-zero density of states at the Fermi level in both configurations.
Additionally, the behavior of the density of states for atoms of the same type within the unit cell is altered due to the influence of the electrodes.
Furthermore, in panel d, it is observed that in the parallel configuration, the monolayer GaTe becomes polarized as a result of its interaction with the ferromagnetic electrodes. In contrast, no such polarization is observed in the antiparallel configuration, as depicted in panel e. This lack of polarization in the antiparallel configuration arises from the opposing spin orientations in the left and right electrodes, which effectively cancel out their individual polarization effects on the barrier.
Moreover, the magnetic moment calculations for the device with parallel configuration confirm a slight polarization of GaTe when it is positioned between two F4GT layers. The magnetic moments of Te and Ga are found to be 0.002  $\mu_{B}$ and 0.004 $\mu_{B}$, respectively. These results suggest that GaTe experiences a subtle magnetic influence in the heterostructure, due to its interaction with the adjacent F4GT layers.

A schematic view of a single layer of GaTe as a spacer between two layers of F4GT, creating a heterostructure of PtTe$_2$/F4GT/GaTe/F4GT/PtTe$_2$ is shown in Fig.~\ref{full_dev}a. The distance between the GaTe layer and left (right) F4GT electrodes was obtained as 3.2 (3.08)~\AA\, which is less than the distance between GaTe layers in AB stacking form (3.81\AA\,).
Figure~\ref{full_dev} indicates the transmission probability for the parallel (P) and antiparallel (AP) spin states in panels (b-d) and (e-g), respectively. In panel (b), a very high spin polarization is observed in the transmission spectrum for the P state at the Fermi level, indicating a preference for one spin orientation over the other. Notably, near the Fermi level, perfect spin filtering occurs, where the transmission is non-zero only for the spin-up channel. This demonstrates that the system acts as an efficient spin filter, allowing only spin-up electrons to pass through the constriction. Upon introducing GaTe as a spacer between F4GT layers, a decrease in the transmission probability is observed, as shown in panel (b) in comparison to Fig.~\ref{TE_bi}b. The $k_\parallel$-resolved transmission probability at the Fermi energy for the spin-up and spin-down channels, presented in panels (c) and (d), respectively, shows that the transmission channels for the spin-down electrons are significantly smaller than those for the spin-up electrons. Both channels exhibit isotropic transmission behavior, with non-zero transmission occurring only at the $\Gamma$ point in the Brillouin zone.

In the AP configuration (panel e), the system exhibits reduced spin polarization and less efficient spin filtering behavior near the Fermi level. The transmission through the device is less sensitive to the electron spin orientation, resulting in a more balanced transmission for both spin-up (panel f) and spin-down (panel g) electrons. The absence of transmission channels in regions away from the $\Gamma$ point indicates that the transmission is confined to a specific momentum range for both spin orientations. Consequently, the transport properties of the system in the AP configuration suggest a less pronounced spin-dependent behavior compared to the P configuration. This observation is consistent with the reported values for P and AP transmission of F3GT/h-BN/F3GT and F3GT/graphene/F3GT heterostructures~\cite{Li2019}. 

Moreover, we focused on the implementation of a bilayer of GaTe as the barrier layer. The total transmission coefficients at the Fermi level for the Fe$_4$GeTe$_2$/bilayer GaTe/Fe$_4$GeTe$_2$ system were determined to be 0.002 and 1.4316e-04 in the parallel and antiparallel magnetic states under zero bias voltage, respectively. Notably, these values are lower than the total transmission coefficients observed when utilizing a monolayer of GaTe as the barrier layer, where they were found to be 0.0250 and 0.0021 for the parallel and antiparallel magnetic states under zero bias voltage, respectively. This outcome highlights the introduction of a bilayer of GaTe as a more effective hindrance to electron transport in the MTJ. Specifically, the bilayer GaTe barrier demonstrates significantly reduced transmittance in both parallel and antiparallel magnetic configurations compared to its monolayer counterpart.

 The total charge current of the device is calculated as the sum of two components: $I=I^\uparrow+I^\downarrow$ (Eq.~\ref{eqLB}). By measuring the currents at different voltages, the TMR ratio can be determined using the formula~\cite{Parkin2004}:
 \begin{equation}
     TMR=\frac{R_{AP}-R_P}{R_{P}}=\frac{1/I_{AP}-1/I_P}{1/I_P}
 \end{equation}
 where $R_P$ and $R_{AP}$ denote the resistances in the P and AP, respectively. Similarly, $I_P (I_{AP})$ represents the total charge currents in the P (AP) magnetization configurations.  
 
\begin{figure}[t]
\centering
\includegraphics[width=0.89\linewidth]{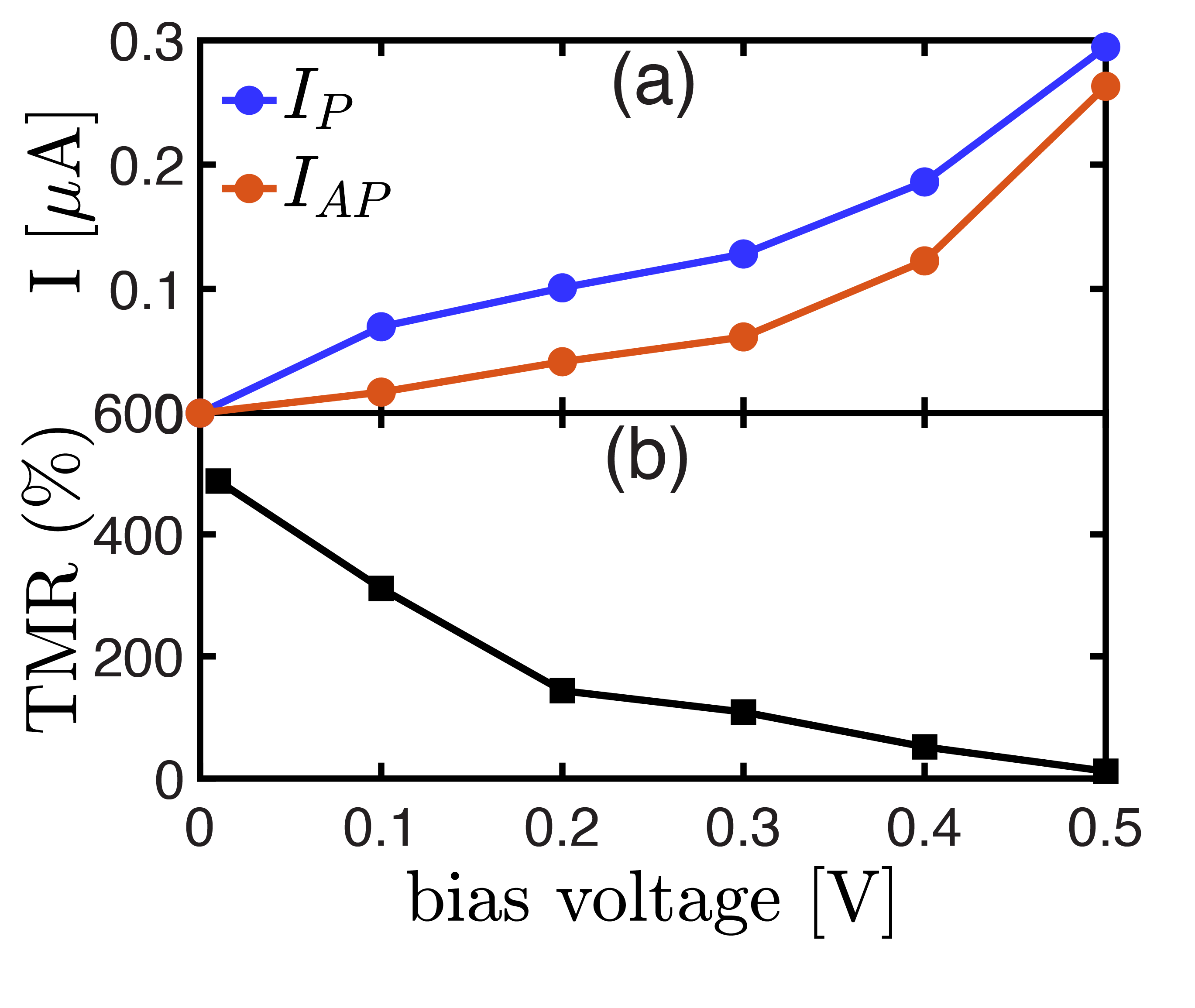}
\caption{\raggedright (a) Total current, a summation of both spin-up and spin-down currents, of Fe$_4$GeTe$_2$/GaTe/Fe$_4$GeTe$_2$ magnetic tunnel junction sandwiched between two PtTe$_2$ electrodes at the parallel and antiparallel state as a function of bias voltage. (b) Variation of the TMR with bias voltage. }
    \label{iv_full}
\end{figure}

 In Fig.~\ref{iv_full}a, the plot illustrates that the current for the P state is higher than the current for the AP state under the given bias voltage. The higher current in the P configuration suggests that electrons with parallel spins have a higher probability of transmitting through the device, leading to more efficient transport of charge carriers. This behavior is consistent with the spin-filtering effect observed in the P state (Fig.~\ref{full_dev}b), where the majority spins experience less resistance and are favored in the transport process. On the other hand, in the AP state, the transmission of both spin-up and spin-down (Fig.~\ref{full_dev}e-g) electrons is more restricted due to the anti-alignment of the magnetic moments. As a result, the total current in the AP configuration is reduced compared to the P configuration.

Tunnel magnetoresistance is a key parameter used to quantify the difference in resistance between the P and AP configurations of an MTJ. A higher TMR indicates a more efficient spin-filtering effect and a larger difference in current between the two spin states, while a lower TMR suggests a reduced difference in current.
In Fig.\ref{iv_full}b, TMR is plotted within a bias range starting from 0.01 V, as it requires a finite bias voltage for accurate representation. The system exhibits a high TMR of 487$\%$ at low bias, surpassing the reported TMR value of 89$\%$ for PtTe$_2$/F4GT/$\alpha$-In$_2$Se$_3$/F3GT/PtTe$_2$ heterostructures~\cite{Su2021}.
However, as the bias voltage increases, the TMR gradually decreases, reaching 12$\%$ at 0.5 V. 
This decline suggests that the efficiency of the spin-filtering effect diminishes under higher voltage conditions. The higher voltage leads to stronger carrier injection in the device, which can modify the spin-dependent transport properties and result in the observed reduction in TMR.
The comparison between the TMR and MR data, as depicted in Fig.~\ref{iv_full}, provides insights. At low bias levels, TMR shows significantly higher values, indicating a robust and efficient magnetoresistive behavior in this voltage range. However, as the bias voltage increases and reaches or exceeds 0.2 eV, the MR begins to exhibit higher values compared to TMR. This observation suggests that the system's magnetoresistive properties vary with the bias voltage, showcasing different behaviors in distinct energy regimes. 
Additionally, the investigation into spin filtering and the potential utilization of other members within the FGT family has produced notable findings. For example, an experiment involving a spin valve device integrated a vertical F3GT/h-BN/F3GT magnetic MTJ with an electrolyte gate, resulting in a MR ratio of 36$\%$ for the intrinsic MTJ. Electrolyte gating further enhanced the TMR ratio of the F3GT/h-BN/F3GT heterostructure from 26$\%$ to 65$\%$~\cite{Zhou} which is lees than the observed value of TMR for F4Gt-based MTJ, specially in low bias voltages. Also, a TMR value of 141$\%$ has been documented in van der Waals magnetic heterostructures comprising F3GT and FePSe$_3$ at 5K~\cite{Huang2023}.
The ability to modulate magnetic fields and magnetoresistance switches presents a promising avenue for controlling the magnetization configuration of the MTJ. In the context of F3GT/Cr$_2$Ge$_2$Te$_6$/F3GT van der Waals junctions, a transition from negative-to-positive magnetoresistance was observed with an increasing bias voltage~\cite{Wang_2023}. This transition, attributed to the changing spin polarizations, was supported by calculated spin-dependent density of states under bias conditions.

\section{Conclusion}
In conclusion, we have investigated the spin-dependent transport properties of $Fe_4GeTe_2/GaTe/Fe_4GeTe_2$ vdW heterostructures sandwiched between PtTe$_2$ electrodes using first-principles calculations and non-equilibrium Green's function method. We analyzed the electronic DOS and MAE of F4GT in both freestanding and device configurations, revealing its ferromagnetic metallic nature and sensitivity to local environments. Through our study, we have demonstrated the formation of spin valves with well-defined spin filtering behavior. Transmission eigenstates of a monolayer F4GT sandwiched between PtTe$_2$ reveal interference patterns influenced by relative phases and localization differing in spin-up and spin-down channels. The transport characteristics of a double-layer F4GT with a ferromagnetic configuration, placed between two PtTe$_2$ electrodes, are found to display remarkable spin polarization of 97$\%$. This indicates a strong tendency for one spin orientation to dominate the transport process. The transport properties of F4GT-based MTJs by introducing GaTe as a spacer between F4GT layers show a remarkable value of 487$\%$ of TMR at low bias surpassing the existing values reported for similar systems in literature. TMR decreases with increasing bias voltage, indicating the modification of spin-dependent transport properties under carrier injection. These findings open up new opportunities for the design and optimization of spintronic devices based on FGT and related heterostructures, advancing the field of spintronics and offering the potential for future technological advancements.

\section{Acknowledgments}
We gratefully acknowledge Prof. Mads Brandbyge of DTU, Denmark for useful discussions. B.S. acknowledges financial support from Swedish Research Council (grant no. 2022-04309) and Carl Tryggers Stiftelse (CTS 20:378). The computations were enabled by resources provided by the National Academic Infrastructure for Supercomputing in Sweden (NAISS) at UPPMAX (NAISS 2023/5-238) and the Swedish National Infrastructure for Computing (SNIC) (SNIC 2022/3-30) at NSC and PDC partially funded by the Swedish Research Council through grant agreements no. 2022-06725 and no. 2018-05973. B.S. also acknowledges the allocation of supercomputing hours in EuroHPC resources in Karolina supercomputer in the Czech Republic.
\bibliography{ref}

\end{document}